\documentclass[times,twocolumn,final]{elsarticle}
\usepackage{cag}
\usepackage{framed,multirow}

\usepackage{siunitx}
\usepackage{booktabs} % \toprule \bottomrule
\usepackage{multirow}

\usepackage{amssymb}
\usepackage{latexsym}
\usepackage{amsmath}
\usepackage[T1]{fontenc}
\usepackage{url}
\usepackage{xcolor}
\definecolor{newcolor}{rgb}{.8,.349,.1}

\usepackage{hyperref}
\hypersetup{
	colorlinks=true, %Colours links instead of ugly boxes
	urlcolor=black, %Colour for external hyperlinks
	linkcolor=cyan, %Colour of internal links
	citecolor=cyan %Colour of citations
}

\usepackage[switch,pagewise]{lineno} %Required by command \linenumbers below

\usepackage{nameref}
\usepackage{wrapfig}
\usepackage{IEEEtrantools}

\usepackage[linesnumbered,lined,boxed]{algorithm2e} % For algorithms

\SetAlFnt{\small}
\SetAlCapFnt{\small}
\SetAlCapNameFnt{\small}
\SetAlCapHSkip{0pt}

\newtheorem{definition}{Definition}

\journal{Computers \& Graphics}

\begin{document}

%\verso{Preprint Submitted for review}
\verso{J. Ma, J. Wang, J. Li, D. Zhang}

\begin{frontmatter}

\title{Real-Time Skeletonization for Sketch-Based Modeling\tnoteref{tnote1}}%
\tnotetext[tnote1]{Only capitalize first
word and proper nouns in the title.}

\author[1]{Jing \snm{Ma}}
%\ead{example@email.com}
    
\author[2]{Jin \snm{Wang}}
%\fntext[fn1]{Footnote 1.}  

\author[2]{Jituo \snm{Li}}
%\fntext[fn1]{Footnote 1.} 

\author[1]{Dongliang \snm{Zhang}\corref{cor1}}
\cortext[cor1]{Corresponding author: Zhejiang University, China}
\emailauthor{dzhang@zju.edu.cn}{Dongliang Zhang}

\address[1]{College of Computer Science, Zhejiang University, Hangzhou, 310058, China}
\address[2]{School of Mechanical Engineering, Zhejiang University, Hangzhou, 310058, China}

%\author[1]{First Author Given Name \snm{Surname}\corref{cor1}}
%\cortext[cor1]{Corresponding author: 
%	Tel.: +0-000-000-0000;  
%	fax: +0-000-000-0000;}
%\emailauthor{example@email.com}{Corresponding Author Name}
%%\ead{example@email.com}
%
%\author[2]{Second Author Given Name \snm{Surname}\fnref{fn1}}
%\fntext[fn1]{Footnote 1.}  
%
%\address[1]{Address, City, Postcode, Country}
%\address[2]{Address, City, Postcode, Country}

%\received{1 February 2017}
\received{\today}
%%%% Do not use the below for submitted manuscripts
%\finalform{28 March 2017}
%\accepted{2 April 2017}
%\availableonline{15 May 2017}
%\communicated{S. Sarkar}

\begin{abstract}
Skeleton creation is an important phase in the character animation pipeline. However, handcrafting skeleton takes extensive labor time and domain knowledge. Automatic skeletonization provides a solution. However, most of the current approaches are far from real-time and lack the flexibility to control the skeleton complexity. In this paper, we present an efficient skeletonization method, which can be seamlessly integrated into the sketch-based modeling process in real-time. The method contains three steps: local sub-skeleton extraction; sub-skeleton connection; and global skeleton refinement. Firstly, the local skeleton is extracted from the processed polygon stroke and forms a subpart along with the sub-mesh. Then, local sub-skeletons are connected according to the intersecting relationships and the modeling sequence of subparts. Lastly, a global refinement method is proposed to give users coarse-to-fine control on the connected skeleton. We demonstrate the effectiveness of our method on a variety of examples created from both novices and professionals.
\end{abstract}


\end{frontmatter}

%\linenumbers

%% Trade names: initial capital letter
\section{Introduction}
\begin{figure*}
	\centering
	\includegraphics[width=\linewidth]{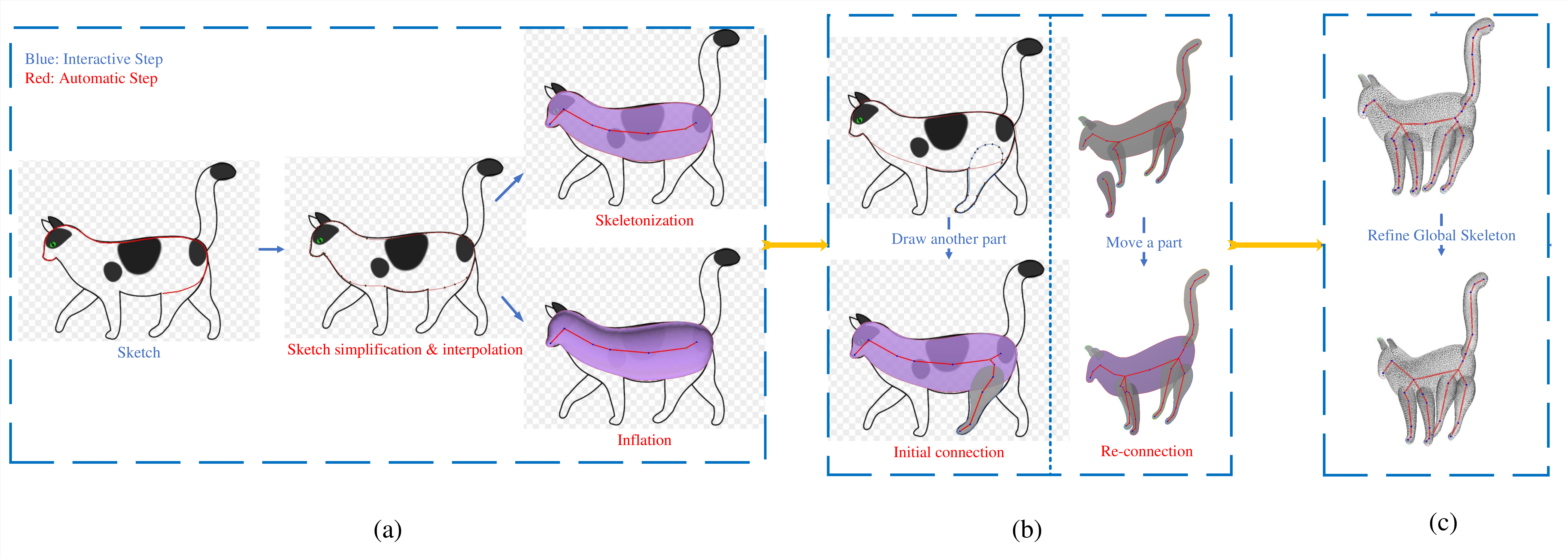}
	\caption{An illustration of creating animatable skeleton using our system: (a) Draw a sketch, a polygon contour is approximated, and a skeleton is extracted from the simplified polygon, meanwhile, the 3D mesh is created through inflation. (b) Draw/move a part, the hierarchical relationship is deducted by fast shape intersection test and modeling sequence, and the precise connecting position from child part to parent part is calculated. (c) Refine the skeleton by multi-level control operations.}
	\label{fig:intro}  
\end{figure*}
There is an increasing need for ready-to-animate models, which takes designers lots of time in shape modeling, skeleton creation, and weight painting. However, manually creating the skeleton is labor-intensive, and requires professional training. Sketch-based tools \cite{teddy_igarashi2006} liberate users from the troublesome of shape modeling, making 3D modeling accessible to novices. There are also lots of work focusing on automatic skeletonization \cite{DistanceTransform_2d3d}\cite{mesh_contraction_au2008}\cite{rosa}\cite{mean_curvature_tagliasacchi2012}\cite{QMat_li2015}\cite{rignet_xu2020}. However, they are computation expensive and far from real-time. The skeleton is extracted only after the entire mesh model is constructed and processed by the algorithm. 

In this paper, we present a real-time skeletonization method for sketch-based modeling, which enables simultaneous shape modeling and skeleton creation. A local skeleton and corresponding mesh are automatically created as soon as the polygon stroke is captured, then the subpart is connected to the existing subpart, finally a coarse-to-fine refinement strategy is provided to the user to control the skeleton complexity, see Fig.~\ref{fig:intro}. Extracting an animatable skeleton in real-time is difficult with three major challenges: $(i)$ how to extract the local skeleton efficiently from an underlying shape; $(ii)$ how to establish the relationship among local skeletons in real-time; and $(iii)$ how to find the optimal skeleton structure suitable for animation.

To solve the first challenge, we construct a straight skeleton \cite{straightskel_aichholzer1996} structure directly from the input polyline and fully utilize Douglas Peucker (DP) simplification \cite{douglas1973algorithms} to accelerate the process. The straight skeleton is extracted by offsetting the polygon edges inward and recording the trace of vertices. We employ straight skeleton as our base skeletonization technique because the computation cost is almost negligible compared with other methods based on binary image \cite{Zhang_thinning}, voxels \cite{lee_3d_thinning}\cite{DistanceTransform_2d3d}\cite{AnimSkelVolNet}, point clouds \cite{rosa}, and meshes \cite{mesh_contraction_au2008}\cite{mean_curvature_tagliasacchi2012}\cite{rignet_xu2020}. The simplicity of polygon data structure makes the algorithm a perfect candidate for immediate skeleton extraction. Our innovation is the full-fledged use of DP technique for speeding up animatable skeleton extraction. As shown in Fig.~\ref{fig:intro}a, we simplify the dense sketch points (red) with DP, the simplified contour not only approximates the shape well but also generates the most concise input for the straight skeleton extraction algorithm. We further propose an efficient BoundedDP algorithm for extracting joints from the axis line of the straight skeleton. Both the axis curvature and silhouette shape are considered, which yields a good result of key joints.

To solve the second challenge, we reuse the skeleton axis and joints extracted from the first step and develop a general cylinder around the axis as well as an inscribed ball around the joint to approximate the 3D shape. This approximation enables the real-time intersection test in the interactive modeling process. Once two 3D shapes are found intersecting with each other, we attach one skeleton to the other according to the modeling sequence. We set the independent subparts as roots, and all follow-up subparts are attached hierarchically according to the intersection relationship and the modeling order. The attached position is calculated from Euclidean distance between parent-skeleton bones and child-skeleton joints.

To solve the third challenge, we equip users with four operations under three level-of-detail controls. As pointed by \cite{rignet_xu2020}, animators need multi-level controls for skeleton complexity, and the fixed number of joints extracted solely from shape topology is not enough to capture user intention.  For example, a hand may be represented by a single medial bone or a finer resolution of hierarchical finger joints. Our four operations provide users flexible control over skeleton complexity. Instead of manually inserting, deleting, and connecting bones, users only need to adjust four parameters to change the skeleton structure. The branch-level operation is based on DP algorithm, which allows users to tune the joints number for a single axis, such as the axis on a finger. For the subpart level and global level, we take inspiration from polygon mesh processing \cite{QEM}\cite{remeshing}\cite{PMP} and skeleton pruning \cite{skeleton_pruning_2012}\cite{skeleton_pruning_2020} since a skeleton can be regarded as an acyclic graph structure (tree). The three operations are \emph{joints merging}, \emph{branch pruning}, and \emph{edge collapsing}. Each operation corresponds to a control threshold that is calculated from the current geometric state of the skeleton. Users can explore different design ideas simply by playing around with these parameters.

In summary, our contributions are four folds: $(i)$ a real-time skeletonization algorithm for immediately constructing an animatable skeleton from user sketch; $(ii)$ an efficient method for fast intersection test and sub-skeletons connection;   $(iii)$ a flexible solution for multi-level of details skeleton control; $(iv)$ an easy-to-use sketch system to create animatable models.
\section{Related Work}
% Straight skeleton
% Mesh contraction, Mean Curvature
% RigMesh, RigNet
%% QMAT, General Cylinder
%% Teddy, Monster Mash
% Head 2
\subsection{Skeletonization}
There are plenty of works for skeletonization. Based on input data types, they can be classified into: binary image and planar curve in 2d; voxels, point cloud, and mesh in 3d \cite{report3d_skeleton2016}. According to the skeletonization technique, they can be classified into propagation-based, and geometric-property-based. In the following, we shall discuss these approaches according to the latter categorization.

\paragraph{Propagation-based methods} The propagation-based methods mimid the grassfire or wavefront transformation moving along the boundary, and the skeleton is the quench where two or more fires/wavefronts meet. In the 2D binary domain, the most representative work is Zhang-Suen Thinning \cite{Zhang_thinning}, which iteratively removes pixels along object borders until no more pixels could be removed. A large body of follow-up works extend this method to 3D voxel domain \cite{lee_3d_thinning}\cite{3d_thinning_1998hybrid}\cite{3d_thinning_1998}\cite{3d_thinning_1999_directional}\cite{3d_thinning_1999}\cite{3d_thinning_2001}\cite{3d_thinning_2002}. These methods are not geometric robust and have the risk of removing important features. Despite the additional processing method \cite{3d_thinning_curve_skel_2008} is proposed to handle excessive removal, the binary propagation approach is generally computation intensive. 
The computation cost is greatly reduced in the 2D geometric domain, namely, the shape represented by 2D lines or curves. Aichholzer introduces a straight skeleton structure \cite{straightskel_aichholzer1996} for the polygon. The structure is progressively constructed by shrinking the polygon and recording the trace of moving vertices. Felkel presents a practical implementation for this algorithm using a doubly-linked list and a priority queue, and the time complexity only subjects to the number of polygon vertices. The 3D counterpart of this algorithm is proposed by \cite{mesh_contraction_au2008}. Similar to shrinking the polygon into zero-area, Au's algorithm guides the triangular mesh shrinking to zero volume by constrained implicit Laplacian smoothing \cite{implicit_smooth_desbrun1999} and defines an edge cost to simplify the collapsed mesh into the curve skeleton. Tagliasacchi et al. \cite{mean_curvature_tagliasacchi2012} further improve Au's algorithm through a detailed analysis of Mean Curvature Flow (MCF) during mesh contraction, and introduce a local remeshing schema to enhance numeric stability. The mesh-based methods are time-consuming incurred by the implicit linear system solving in each contraction iteration, and the generated curve skeleton needs post-processings to be animatable. Compared with a curve skeleton, the straight skeleton structure \cite{straightskel_aichholzer1996} conforms more to an animatable skeleton. Our skeletonization algorithm benefits from such conformance and the simpleness of the planar polygon. Meanwhile, it does not need an expensive mesh contraction process to create the 3D skeleton.

\paragraph{Geometric property-based methods} The geometric methods strive to find the shape centers by analyzing the translation and rotation property of the inner region towards the boundary. One popular choice is Medial Axis Transformation (MAT). The MAT skeleton is defined as the locus of the centers of all maximally inscribed circles (in the 2D domain) or spheres (in the 3D domain). Plenty of MAT literature exist for different geometric
entities: the 2D/3D binary ones \cite{zhou1999efficient}\cite{bitter2001penalized}\cite{mat2003skeleton}\cite{hassouna2005robust}\cite{DistanceTransform_2d3d}, the planar curve \cite{MAT_spline}, the point clouds \cite{MAT_point_clouds}, and the mesh \cite{QMat_li2015}. These methods are computationally intensive, numerically unstable excepting the work for planar curve \cite{MAT_spline}, and are sensitive to boundary perturbations. In addition to the distance transformation property, rotation traits are used in \cite{rosa}. The curve skeleton is discovered by finding a rotational symmetry axis (ROSA) for an oriented point cloud. They propose an iterative planar cuts method to find the optimal ROSA plane. This method is extended by \cite{general_cylinder_zhou2015} to the mesh. They use the ROSA method to find a general cylinder decomposition and partial curve skeleton given a polygon mesh. However, these methods are also time-consuming, and the extracted curve skeleton is not ready for animation.

\subsection{Automatic Rigging}
Automatic rigging creates a ready-to-animate bone skeleton \cite{mesh_contraction_au2008} and binds the skeleton to the mesh. The pioneering work of automatic rigging is proposed by \cite{automatic_rigging_baran2007}, which fits a predefined skeleton template to the mesh, and calculates the skinning weight through heat diffusion. However, their method is limited by the skeleton template and fails to cope with various shapes. On the contrary, our work is highly adaptable to various shapes in a creating-on-the-go fashion. It is closely related to RigMesh proposed by \cite{rigmesh_borosan2012}, in which shape modeling and skeleton creating are handled simultaneously. RigMesh uses Constrained Delaunay Triangulation to decompose the silhouette into needle-like triangles and constructs the skeleton by connecting triangle centers. It defines three strategies for skeleton connection: splitting, snapping, and connecting. When a  user drags one part close to the other, a connecting suggestion from these three cases is made. The underlying meshes are merged immediately after the user accepts the recommendation. This strategy is unintuitive and inaccurate,  leading to the attached mesh deviating from its original position. In contrast, our algorithm can calculate the exact attach position conforming to the user's original intention. With the rapid development of deep learning techniques, the recent work uses a deep neural network to solve the automatic rigging problem \cite{rignet_xu2020}, however, their approach is limited to the training set and does not work well for arbitrary shapes, and the method also has a high requirement for devices. Our algorithm works for the novel shapes that a network is never trained before, and it is also practical to be deployed in low-end devices.

\subsection{Sketch Modeling}
Sketch-based system pioneered by Igarashi et al. \cite{teddy_igarashi2006} greatly reduces the workload of shape modeling. Users only need to draw few strokes to create a 3D shape. And the subsequent works \cite{tai2004prototype}\cite{cherlin2005sketch}\cite{nealen2005sketch}\cite{schmidt2007shapeshop}\cite{alexe2007shape}\cite{nealen2007fibermesh}\cite{sugihara2008sketch}\cite{cordier2011sketching} make sketch modeling more complete, with the power of creating complex shapes. The recent work of MonsterMash  \cite{monster_mash_dvorovzvnak2020} propose a framework for casual sketch modeling and animation prototyping under single-view, the work provides lots of fun to novice users, however, the strength of simple interaction is also a weakness of the system, especially for professional users who want an accurate model. Our system allows users to create accurate shapes under different views without losing the interaction simplicity.

\section{Our Method}
\subsection{Local Sub-Skeleton Extraction} \label{sec:local_skel_extract}
\begin{figure*}
	\centering
	\includegraphics[width=\linewidth]{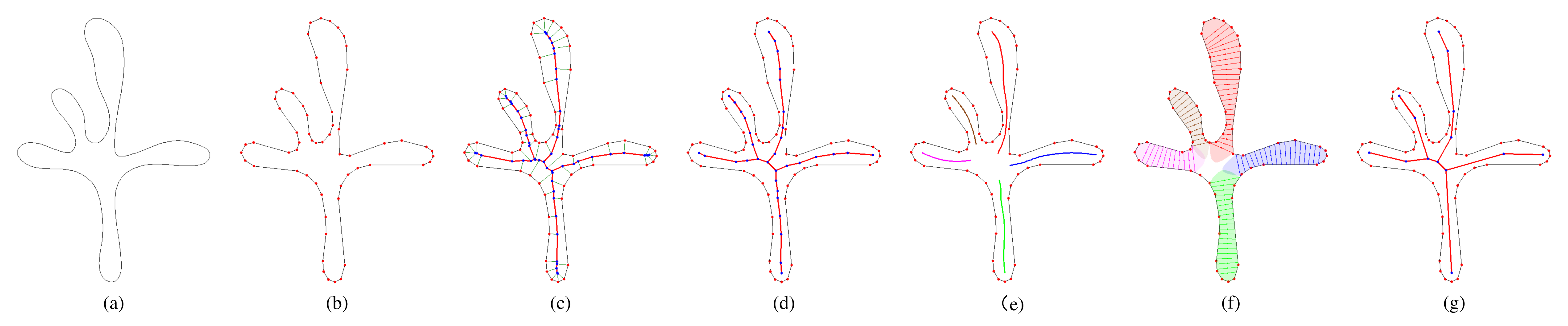}
	\caption{Local sub-skeleton extraction.  (a) Processed sketch line. (b) Simplified polygon. (c) Straight skeleton traced by propagating polygon edges inward. (d) Animatable clean skeleton by removing redudant vertices and edges. (e) Spline interpolated by long polyline branch. (f) Slice  uniformly along the spline to capture contour variation. (e) Simplified skeleton by bounded DP simplification.}
	\label{fig:sub_skel_extract}
\end{figure*}
\begin{figure}[t]
	\centering
	\includegraphics[width=0.95\linewidth]{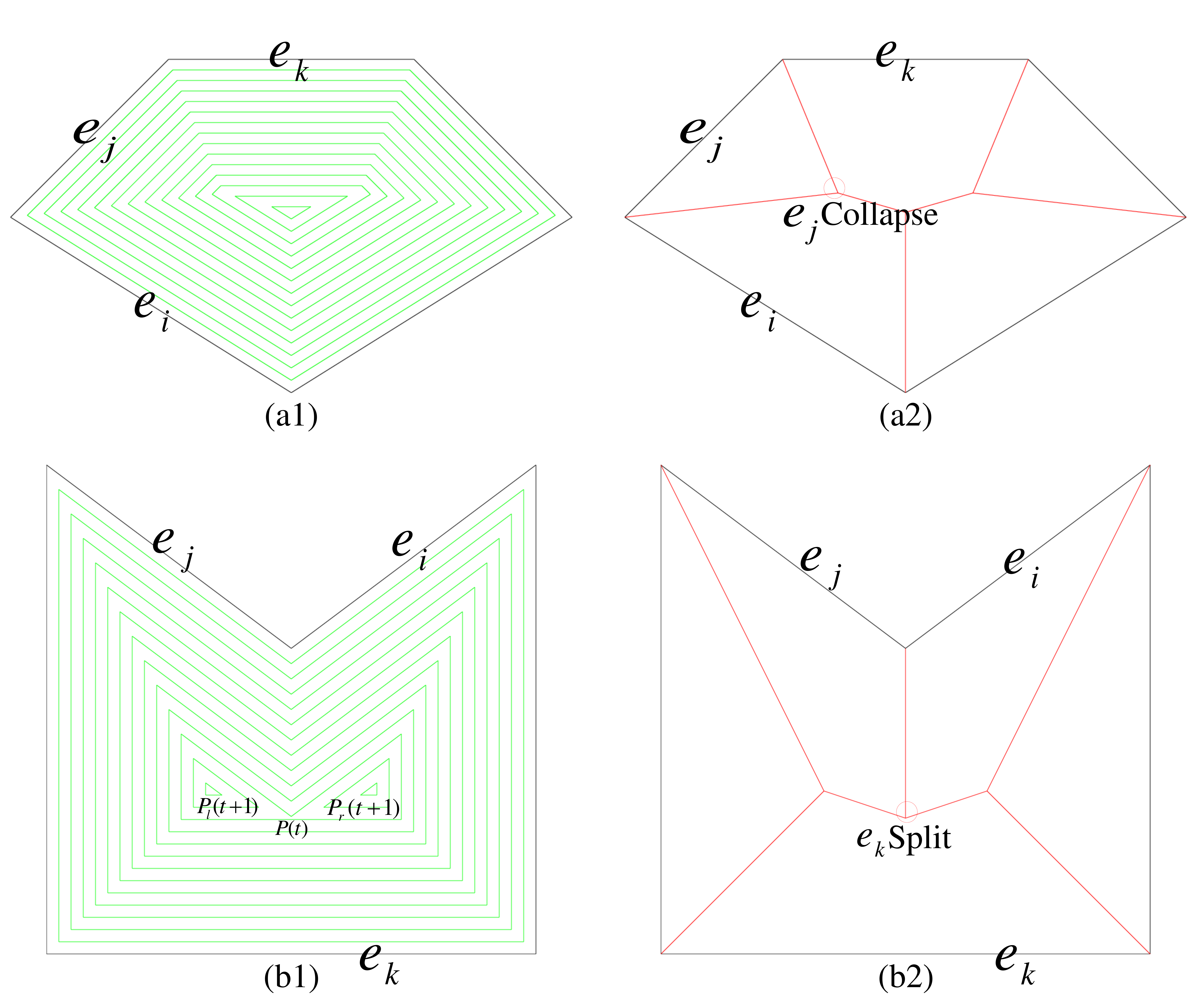}
	\caption{Black: the initial silhouette polygon. Green: the intermediate offsetting polygons. Red: the traced straight skeleton. (a) Edge Event: non-consecutive edges $e_i$ and $e_k$ collide, and $e_j$ collapses. (b) Split Event: consecutive edges $e_i$ and $e_j$ collide simultaneously with opposite edge $e_k$, and $e_k$ splits. The polygon $P(t)$ at the time $t$ splits into two smaller polygons $P_l(t+1)$ and $P_r(t+1)$ at the next time $t+1$.}
	\label{fig:events}
\end{figure}
\begin{figure}[h]
	\centering
	\includegraphics[width=0.5\linewidth]{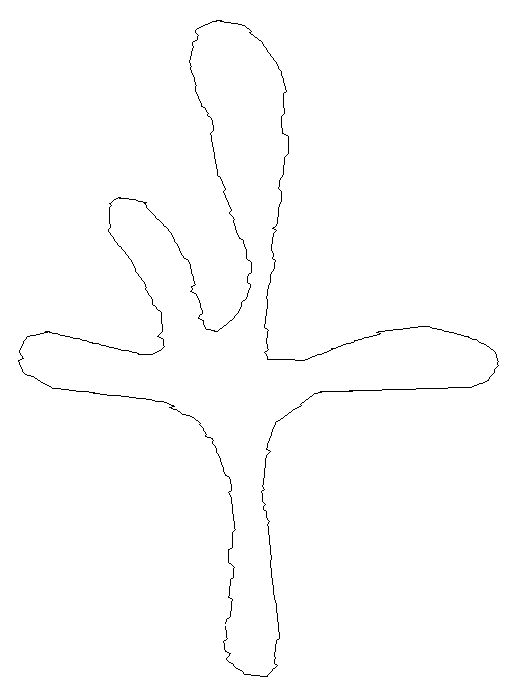}
	\caption{Unsteady stroke. In our system, the perimeter of the stroke is 1972.28, and the step size for uniform discretization is 10.}
	\label{fig:rawsketchnoised}
\end{figure}

Previous works extract skeleton from the merged mesh \cite{mesh_contraction_au2008}\cite{mean_curvature_tagliasacchi2012}, which is slow in interactive scenarios. To achieve real-time performance, we generate the local sub-skeleton directly from stroke-input. The local method contains three sub-steps: $(i)$ simple polygon acquisition; $(ii)$ straight-skeleton extraction; $(iii)$ and straight-skeleton simplification, shown in Fig.~\ref{fig:sub_skel_extract}. In this section, we first introduce the definition of simple polygon and straight skeleton. Then, we describe the details of three sub-steps.
\begin{wrapfigure}{R}{0.2\linewidth}
	\centering
	\includegraphics[width=\linewidth]{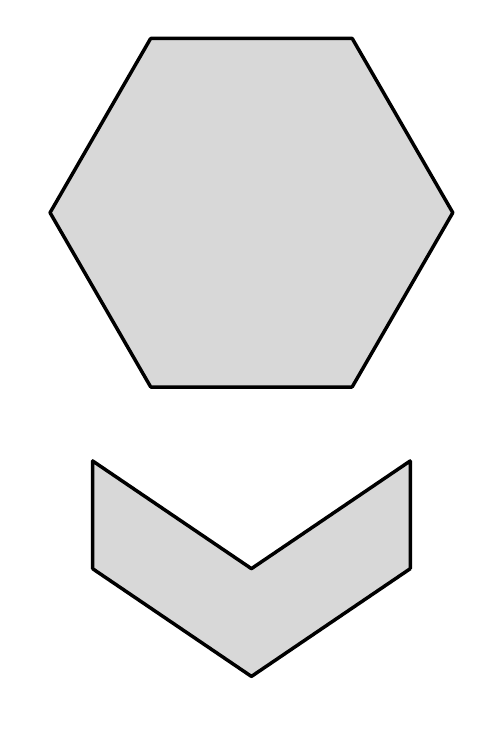}
	\caption{Simple polygon}\label{fig:simplepolygon}
\end{wrapfigure}
\begin{definition}[Simple Polygon]
	A simple polygon is a polygon that does not intersect itself and has no holes. See Fig.~\ref{fig:simplepolygon} for an illustration.
\end{definition}
\begin{definition}[Straight Skeleton]
	A straight skeleton is a special partitioning of a polygon into monotone regions traced by a continuous inward offsetting of the contour edges \cite{straightskel_aichholzer1996}. See Fig.~\ref{fig:events} for an illustration.
\end{definition}
\paragraph{Simple polygon acquisition} 
To acquire valid and concise input for the local skeletonization algorithm, the raw sketch line needs to be processed. The sketch line is usually smooth but sometimes may contain noises, shown in Fig.~\ref{fig:rawsketchnoised}. We smooth out these noises by uniformly discretizing the raw line by a small length, which yields a smooth input for our algorithm, shown in Fig.~ \ref{fig:sub_skel_extract}a. Then we apply Douglas-Peucker (DP) algorithm \cite{douglas1973algorithms} to find a simplified polygon best approximating the shape, shown in Fig.~\ref{fig:sub_skel_extract}b, details of DP are introduced in  \emph{Section \nameref{para:ss_simp}}. Lastly, we arrange the polygon vertices as counter-clockwise oriented to be ready for the next step. We denote the simple polygon as $\mathcal{P}(\mathcal{V}, \mathcal{E})$ consisting a set of vertices $\mathcal{V}=\{v_i\}$ and edges $\mathcal{E}=\{e_i\}$.

\begin{figure}
	\centering
	\includegraphics[width=\linewidth]{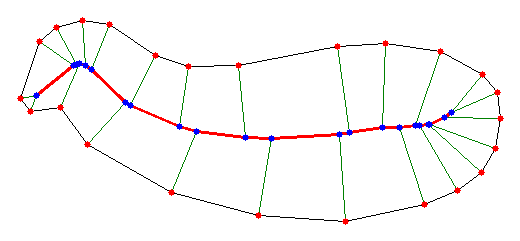}
	\caption{Straight skeleton. A straight skeleton contains two types of vertices: border vertices (red circle) and skeleton vertices (blue cirlce), and three types of edges: border edges (black segment), peripheral edges (green segment), and skeleton edges (red segment).}
	\label{fig:straight_skeleton}
\end{figure}

\paragraph{Straight skeleton extraction} 
Then a straight skeleton $\mathcal{S(\mathcal{P})}$ is extracted from the simple polygon $\mathcal{P}(\mathcal{V}, \mathcal{E})$.  The basic idea is to propagate edges inward like setting fire around the border, and the straight skeleton is traced by bisectors of polygon edges. The polygons here refer to the initial silhouette polygon and intermediate offsetting polygons generated by propagation, see Fig.~\ref{fig:events}. In a formal description, all edges move at the same speed along their respective perpendicular direction. During propagation, two types of events may changes polygon's topology: $(i)$ the edges $e_{i}$ and $e_{k}$ (adjacent to the edge $e_j$) collide, and $e_j$ vanishes; $(ii)$ an edge $e_k$ collides with two consecutive edges $e_i$ and $e_{j}$, splitting $e_k$ into two edges on collision position. We define the former as \emph{Edge Event} and the latter as \emph{Split Event} following the same convention from \cite{sske_impl_felkel1998} and \cite{sskel_impl_cgal_cacciola2004}, see Fig.~\ref{fig:events}. The method is summarized in Algorithm~\ref{alg:one}.
\paragraph{Straight-skeleton simplification} \label{para:ss_simp}
\begin{wrapfigure}{R}{0.2\linewidth}
	\centering
	\includegraphics[width=\linewidth]{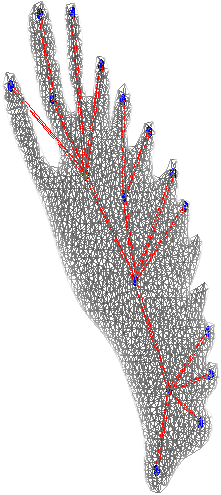}
	\caption{Over simplification}\label{fig:out_of_shape}
\end{wrapfigure}
The straight skeleton from the previous step contains many unnecessary vertices and edges as shown in Fig.~\ref{fig:sub_skel_extract}c and Fig.~\ref{fig:straight_skeleton}. Skeleton vertices traced from high curvature borderline are clustering together. Neither the peripheral edges nor the very short skeleton edges are wanted for constructing an animatable skeleton. Thus, we remove all peripheral edges, and collapse short skeleton edges less than a given threshold, shown in Fig.~\ref{fig:sub_skel_extract}d. In our experiment, we set the collapse threshold as $0.5*average(skeleton\;edge\;length)$, which produces a clean animatable skeleton in most cases. Still, this skeleton is not concise enough. Designers may find that nodes along the long axis are superfluous. The Douglas Peucker algorithm \cite{douglas1973algorithms} is a good option to do the simplification. However, only considering the axis curve while ignoring the shape will lead to an over-simplified or even out-of-shape skeleton, see Fig.~\ref{fig:out_of_shape}. To address this problem, we propose a BoundedDP algorithm considering both the axis and the shape. Details of this algorithm will be introduced in the following.

A spline curve is interpolated for each long polyline, shown in Fig.~\ref{fig:sub_skel_extract}e. We observe that the region enclosing a long branch usually forms a general cylinder, which could be used as boundary restriction during simplification. To extract the general cylinder, we first generate a set of uniformly sampled points ${\{p_i\}}$ along the axis curve, and compute a segment $s_i$ at each $p_i$ by intersecting the polygon with an infinite line perpendicular to the central axis at $p_i$, see Fig.~\ref{fig:sub_skel_extract}f. Meanwhile, the intersected edges on the polygon are also recorded, forming two nearly parallel lines on the left and the right side of the central axis. By closing open holes at the end of these two lines, we get the general cylinder region $\Omega$, see shadow area in Fig.~\ref{fig:sub_skel_extract}f. Then $p_i$ and $\Omega$ are used as input for BoundedDP to get the simplified polyline $l(\{p_i\}, \{e_i\})$ with each polyline edge $e_i$ lying inside the bounded region $\Omega$. 
\begin{figure*}
	\centering
	\includegraphics[width=\linewidth]{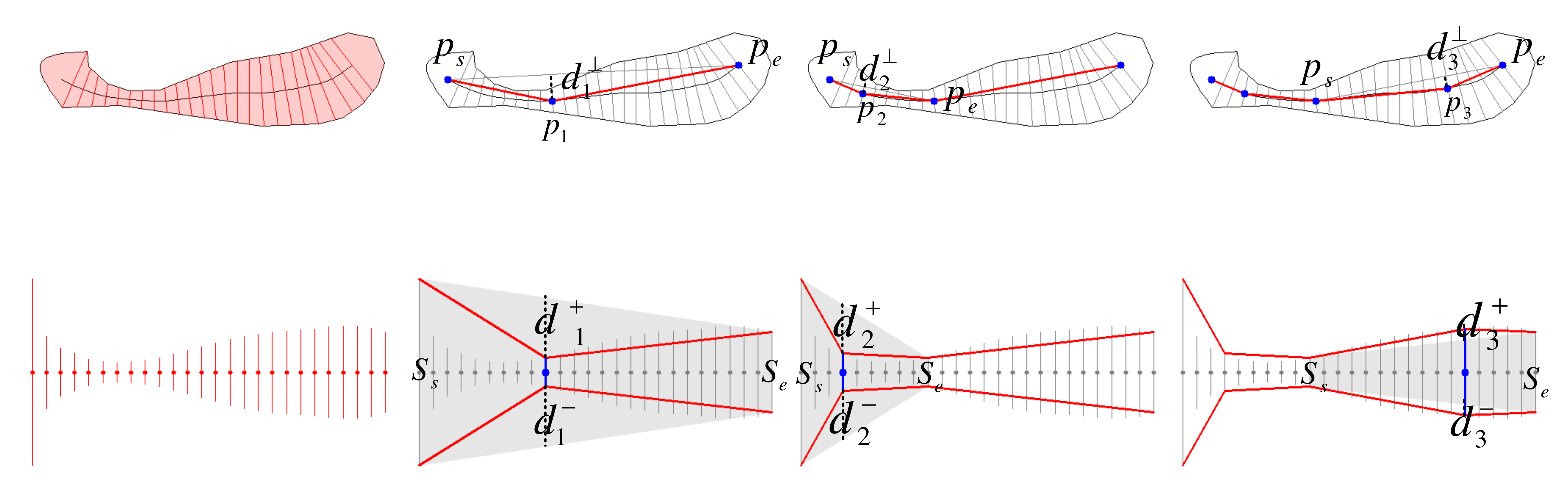}
	\caption{BoundedDP simplification (red branch in Fig.~\ref{fig:sub_skel_extract}f). The upper part of the first column shows an axis curve and its enclosing general cylinder. The lower part of the first column shows a parallel arrangement of intersected segments, which are acquired by uniformly slicing along the axis. The next three columns show the first three steps of BoundedDP. Better approximations are found for both the line and the shape as the algorithm progresses.}
	\label{fig:boundeddp}
\end{figure*}

The original Douglas Peucker algorithm works in a recursive greedy manner: at each step, it selects a point $p_i$ with maximum distance to the line $p_s p_e$. If the distance $d_i^{\perp}$ is larger than a threshold $\epsilon$, $p_i$ is regarded as important for fitting the original line, and is added to the output, see the first row of Fig.~\ref{fig:boundeddp}. The algorithm then recursively calls itself on the split line, $p_s p_i$ and $p_i p_e$, finding the furthest point in each range. The recursion terminates until no point can be found with its distance larger than the threshold $\epsilon$. Two essential ingredients contribute to a good approximation: (i) \emph{the criteria} to select the important point; (ii) \emph{the threshold} to balance between simplicity and approximation quality. For \emph{the criteria}, instead of only considering the variation of axis curve, our algorithm also takes shape variation into account, which leads us to the following formula:
\begin{IEEEeqnarray}{rCl}
E_i & = & E_{p_i} + \alpha_{s} E_{s_i} \\
& = & d_i^{\perp} + \alpha_{s} (d_i^{+} + d_i^{-})
\end{IEEEeqnarray}
where $E_{p_i}$ measures the axis variation, namely, the perpendicular distance $d_i^{\perp}$ from current point $p_i$ to the straight line connecting two endpoints, see the first row of Fig.~\ref{fig:boundeddp}. To capture the shape variation, we add the second term $E_{s_i}$ to the formula, measuring the distances $d_i^{+}$ and $d_i^{-}$ from two endpoints of segment $s_i$ to the two sides of the trapezoid, which is bounded by two end segments $s_s$ and $s_e$,  see the dotted line (distances) and the shaded area (trapezoid) of the second row in Fig.~\ref{fig:boundeddp}. Notice we rearrange the segments $\{s_i\}$ parallelly by stretching the axis curve to a horizontal straight line, filtering the axis variation out of the second term. In each step, we choose the point maximizing the error $E_i$, which is a summation of axis variation and a weighted ($\alpha_s$) shape variation. The point selection procedure is summarized in Algorithm~ \ref{alg:point_sel}. For \emph{threshold}, the larger the value is, the more simplified result we get, sometimes inducing undesirable skeleton, see Fig.~\ref{fig:out_of_shape}. To select a proper $\epsilon$, we initialize the threshold with a fairly large value $\epsilon_0$, and try to find an approximate polyline $l(\{p_i\}, \{e_i\})$. Then we check whether all edges $e_i$ of the simplified polyline are inside the shape $\Omega$. If there exist intersections between $e_i$ and the shape boundary $\partial \Omega$, we tune down the current threshold $\epsilon$ by a factor of $0.8$. This procedure is summarized in Algorithm \ref{alg:thresh_adj}.
\IncMargin{1ex}
\begin{algorithm}[t]
	%\SetAlgoNoLine
	\SetKwFunction{BoundedDP}{BoundedDP}
	\SetKwFunction{PerpendicularDistance}{PerpendicularDistance}
	
	\KwIn{$\mathbf{p}=\{p_i\}$, $\mathbf{s}=\{s_i\}$, start index $i_{st}$, end index $i_{en}$, threshold $\epsilon$}
	\KwOut{Array of booleans $\mathbf{b}=\{b_i\}$ marking which $p_i$ to be retained.}
	\BlankLine
	\lIf{$i_{en} \le i_{st}+1$}{return}
	\tcp{Find point with max error}
	$i_{max} \leftarrow i_{st}$\;
	$E_{max} \leftarrow 0$\;
	\For{$i \leftarrow i_{st}+1$ \KwTo $i_{en}$}{
		$E_{p_i} \leftarrow$ \PerpendicularDistance{$p_i$, Line($p_{i_{st}}$, $p_{i_{en}}$)}\;
		$E_{s_i} \leftarrow d_i^+ + d_i^-$ \tcp*[r]{shape variation, see Fig.\ref{fig:boundeddp}} %ToDO
		$E_i \leftarrow E_{p_i} + E_{s_i}$\;
		\If{$E_i > E_{max}$}{
			$i_{max} \leftarrow i$\;
			$E_{max} \leftarrow E_i$
		}
	}
    $b[i_{max}] \leftarrow true$\;
    \tcp{Recursive selection}
	\If{$E_{max} > \epsilon$}
	{
		\BoundedDP{$\mathbf{p}$, $\mathbf{s}$, $i_{st}$, $i_{max}$, $\mathbf{b}$}\;
		\BoundedDP{$\mathbf{p}$, $\mathbf{s}$, $i_{max}$, $i_{en}$, $\mathbf{b}$}\;
	}
	\caption{BoundedDP: point selection.}\label{alg:point_sel}
\end{algorithm}\DecMargin{1ex}
\IncMargin{1ex}
\begin{algorithm}[t]
	%\SetAlgoNoLine
	\SetKwFunction{BoundedDP}{BoundedDP}
	\SetKwFunction{IsInsideShape}{IsInsideShape}
	
	\KwIn{$n$ points $\mathbf{p}=\{p_i\}$ and perpendicular segments $\mathbf{s}=\{s_i\}$ uniformally sampled along axis curve, bounded region $\Omega$, a relatively large initial threshold $\epsilon_{0}$, adaptive ratio $\alpha \in (0, 1)$}
	\KwOut{Simplified polyline $l(\{p_i\}, \{e_i\})$}
	\BlankLine
	$\epsilon \leftarrow \epsilon_{0}$ \tcp*[r]{initial threshold}
	\While{not \IsInsideShape{$l$, $\Omega$}}{
		$\mathbf{b} \leftarrow$ $\{false\}$\; 
		\BoundedDP{$\mathbf{p}$, $\mathbf{s}$, $1$, $n$, $\mathbf{b}$}\;
		clear previous $l$, add $p_i$ to $l$ if $b_i$ is $true$\;
		$\epsilon \leftarrow \alpha * \epsilon$ \tcp*[r]{reduce threshold}
	}
	\caption{BoundedDP: threshold tuning.}\label{alg:thresh_adj}
\end{algorithm}\DecMargin{1ex}

\subsection{Sub-Skeleton Connection}
Before starting this section, we give the following definition:
\begin{definition}[Junction joint]
	Joint adjacent to three or or more bones. 
\end{definition}
\begin{definition}[Sleeve joint]
	Joint adjacent to two bones.
\end{definition}
\begin{definition}[Terminal joint]
	Joint only adjacent to one bone.
\end{definition}
\begin{definition}[Branch]
	A branch begins with either a junction joint or terminal joint, and ends with a junction joint or terminal joint, probably with sleeve joints between two ends. 
\end{definition}

\begin{figure}
	\centering
	\includegraphics[width=0.95\linewidth]{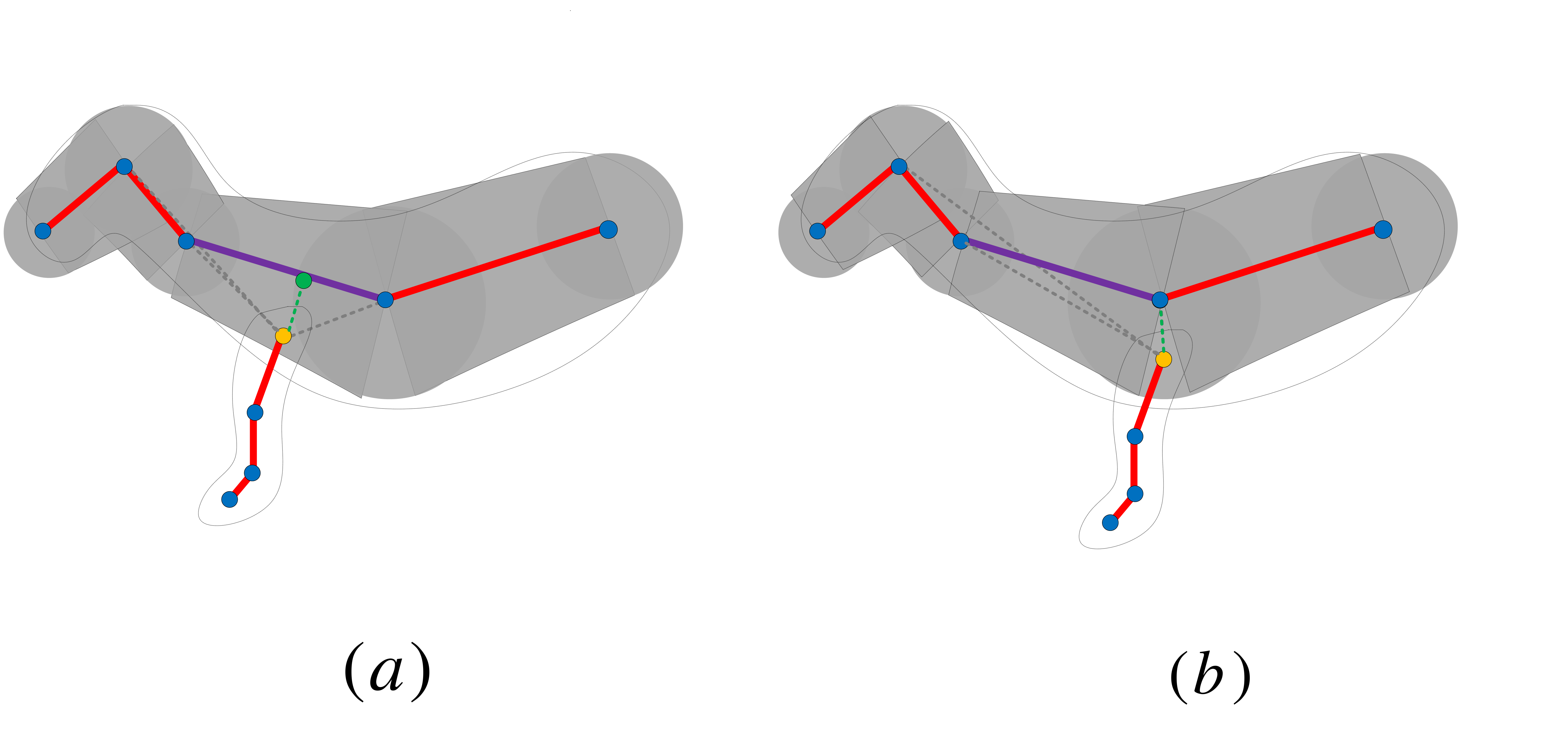}
	\caption{Sub-skeleton connection: connecting child-leg-skeleton to parent-torso-skeleton. The shadowed area shows the cylinder-ball approximation of the shape, which enables real-time intersection test. The dotted line shows the shortest distance from the candidate child joint to each parent bone. (a) Bone Split Event. (b) Joint Connect Event.}
	\label{fig:boneevent}
\end{figure}

We connect sub-skeletons in an interactive pairwise fashion. This conforms to the real scenarios in which users interactively create 3D models part by part. The new subpart is connected to the existing subpart, meanwhile, the new sub-skeleton is connected to the corresponding sub-skeleton. When the user creates a new subpart or moves an existing subpart, see Fig.~\ref{fig:intro}b, we instantly check whether the current subpart intersects with other subparts. If the current subpart intersects with another subpart, we connect the skeleton in the current subpart to the skeleton in the intersected subpart. To make intersection test real-time, we reuse the local skeleton extracted from the previous step. Specifically, we find a cylinder-ball approximation for the mesh shape, see Fig.~\ref{fig:boneevent}. Each skeleton edge serves as the medial axis of the general cylinder, and each bone serves as the center of the inscribed ball. The ball’s radius is the average distance from the joint to the cross-section contour, which is precomputed by slicing the mesh along the bone's perpendicular direction. Once the radius of the two end joints is computed, the radius for the general cylinder can be calculated by linear interpolation of the two end radius. We denote sub-skeleton on the current subpart as child-skeleton and sub-skeleton on the previous intersected subpart as parent-skeleton. Initially, we set the first subpart as the root. If a subpart does not intersect with any previous subpart, we also add it to the roots set. All subparts created after and intersected with a root becomes the root's child. All subparts created after the child and intersected with the child become the child’s child (one of the root’s descendants), so on and so forth, formulating a tree hierarchy. If the terminal joints or junction joints of the child skeleton lies inside the cylinder-ball region of the parent, we regard that the two parts intersect with each other and start locating a more precise position from the child-skeleton to the parent-skeleton for attaching.

\begin{figure}
	\centering
	\includegraphics[width=0.98\linewidth]{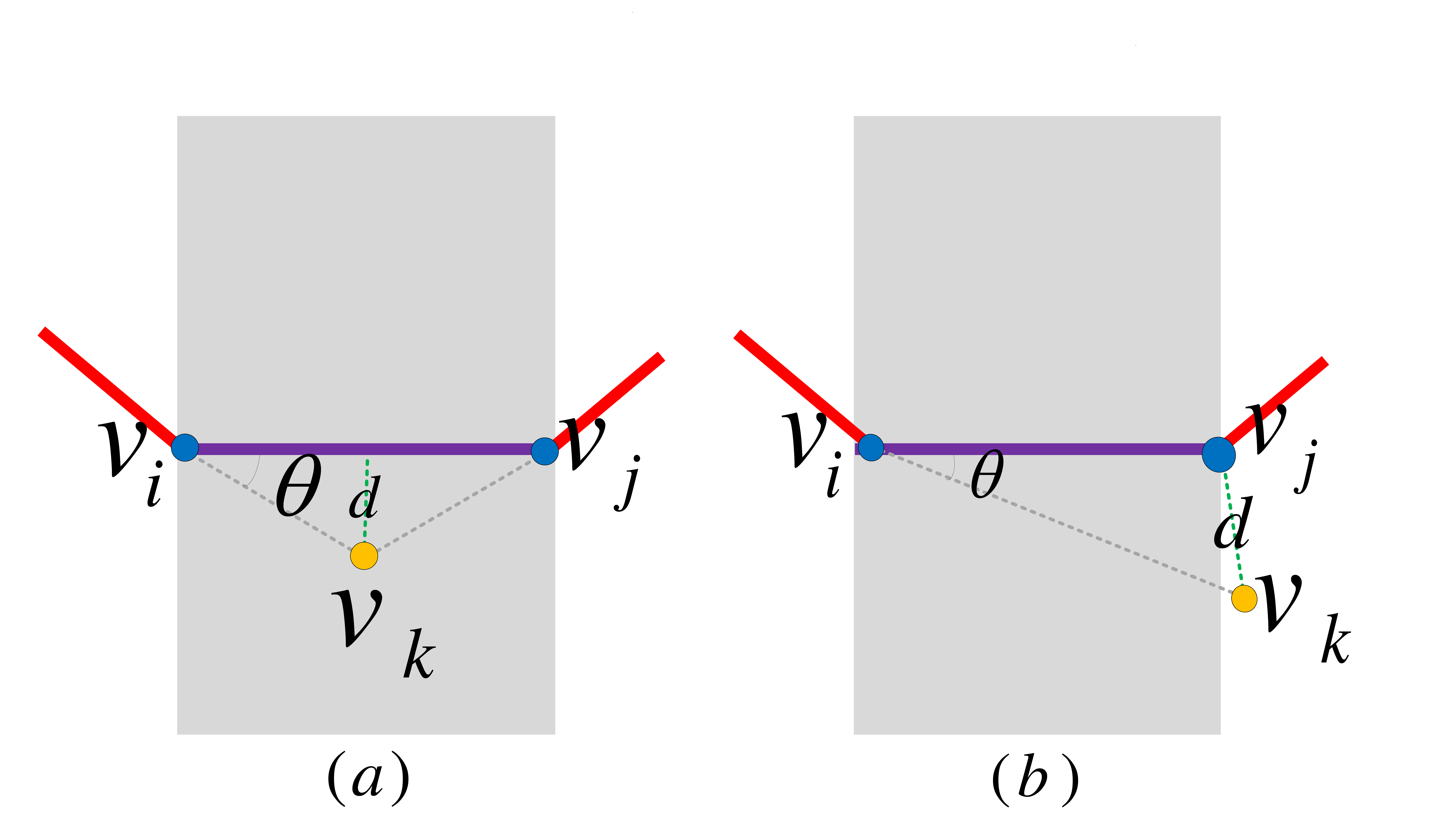}
	\caption{Distance from child joint (yellow) to parent bone (purple).}
	\label{fig:boneconnecteq}
\end{figure}
On child-skeleton, terminal joints and junction joints are selected as the candidates to be connected to the parent bones.We then tries to find an optimal position on the parent-skeleton where the child's candidate joints could be connected to, the one minimizing the distance between the child joint and the parent-skeleton. Firstly, we search for the best bone/joint from the parent skeleton. We compute the shortest distance from the child candidate joint to each parent bone. The bone with minimum distance is selected as the best spot for connection, see the purple segment in Fig.~\ref{fig:boneevent}. Given a parent bone with start joint $v_i$ and end joint $v_j$, the shortest distance from parent's bone to child's joint $v_k$ is calculated as:
\begin{equation} \label{eq:bone_conn}
d = 
\begin{cases}
|\overrightarrow{v_k v_i}|\sin{\theta} & \text{if } \overrightarrow{v_i v_j} \cdot \overrightarrow{v_i v_k} \ge 0 \text{ and }  \overrightarrow{v_j v_k} \cdot \overrightarrow{v_j v_i} \ge 0,\\
min\{|\overrightarrow{v_k v_i}|, |\overrightarrow{v_k v_j}|\} & \text{otherwise}
\end{cases}
\end{equation}
where $\overrightarrow{v_i v_j} \cdot \overrightarrow{v_i v_k} \ge 0 \text{ and }  \overrightarrow{v_j v_k} \cdot \overrightarrow{v_j v_i} \ge 0$ defines an influence region for each bone, see shadowed area in Fig.~\ref{fig:boneconnecteq} and Fig.~ \ref{fig:boneevent}. The influence region is bounded by two  planes perpendicular to the bone. Only when the child joint lies inside the two planes, the perpendicular point-to-line distance is used Fig.~\ref{fig:boneconnecteq}a. If child joint $v_k$ is outside the bone's influence region, we calculate the Euclidean Distance from the joint to the bone's two end joints respectively, and choose the smaller one as the shortest distance Fig.~\ref{fig:boneconnecteq}b.

Then, we connect child-skeleton to parent-skeleton by choosing the closest joint-joint or joint-bone pair. Two types of events occur: \emph{bone split event} and \emph{joint connect event}, see Fig.~\ref{fig:boneevent}. \emph{Bone split event} occurs when the child‘s joint lies inside the parent bone's influence region. We create a new joint on the parent bone, green circle in Fig.~ \ref{fig:boneevent}a, and split the bone into two. Then a new bone is created between the new joint and the child joint. Notice if the child’s joint is close to the parent bone's end, one of the split bones will be very short. We address this problem in Sec\ref{sec:GlobalRefine}. \emph{Joint connect event} occurs when the child’s joint lies outside the parent bone's influence region.  We create a new bone connecting the child joint and the nearest parent joint.
\subsection{Global Skeleton Refinement} \label{sec:GlobalRefine}
\begin{figure*}
	\centering
	\includegraphics[width=\linewidth]{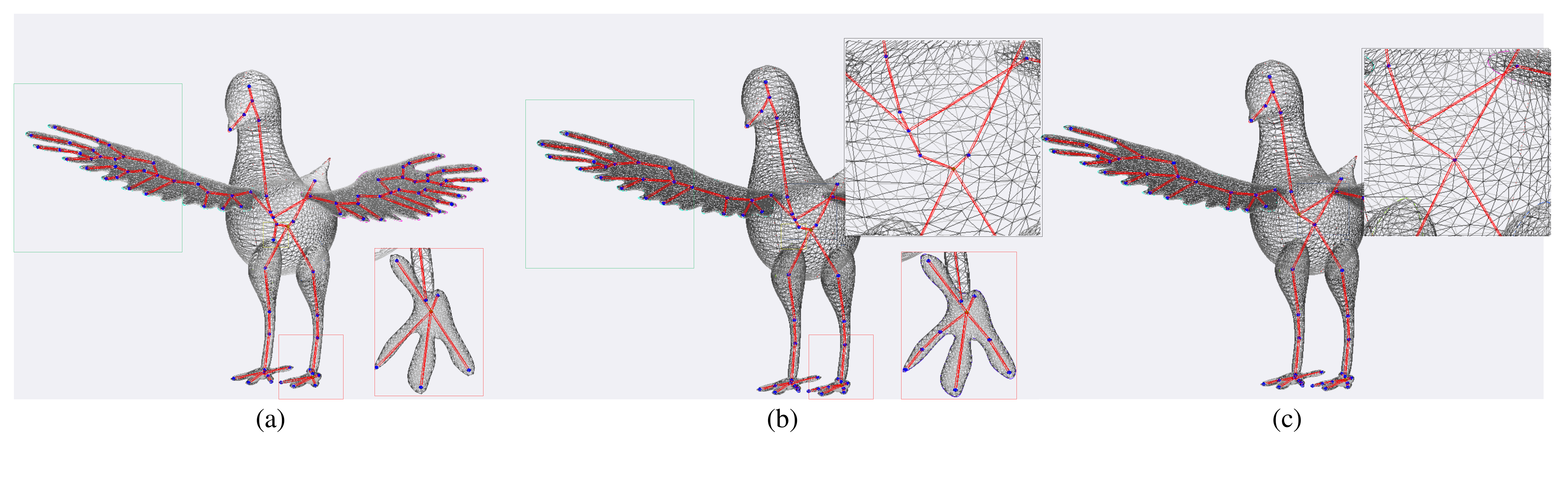}
	\caption{Refine a bird skeleton. (a) The initial skeleton under default parameters, constructed by our system automatically. Green boxes (a)-(b): simplify branches on wings by increasing threshold $\epsilon_{s}$. Red boxes (a)-(b): acquire more complex structure on claw by reducing threshold $\epsilon_{s}$. Yellow boxes (a)-(b): trim short JT branch by increasing threshold $\epsilon_t$. Black boxes (b)-(c): collapse short internal bones by increasing threshold $\epsilon_c$.}. 
	\label{fig:refine_bird}
\end{figure*}
\begin{figure}
	\centering
	\includegraphics[width=0.95\linewidth]{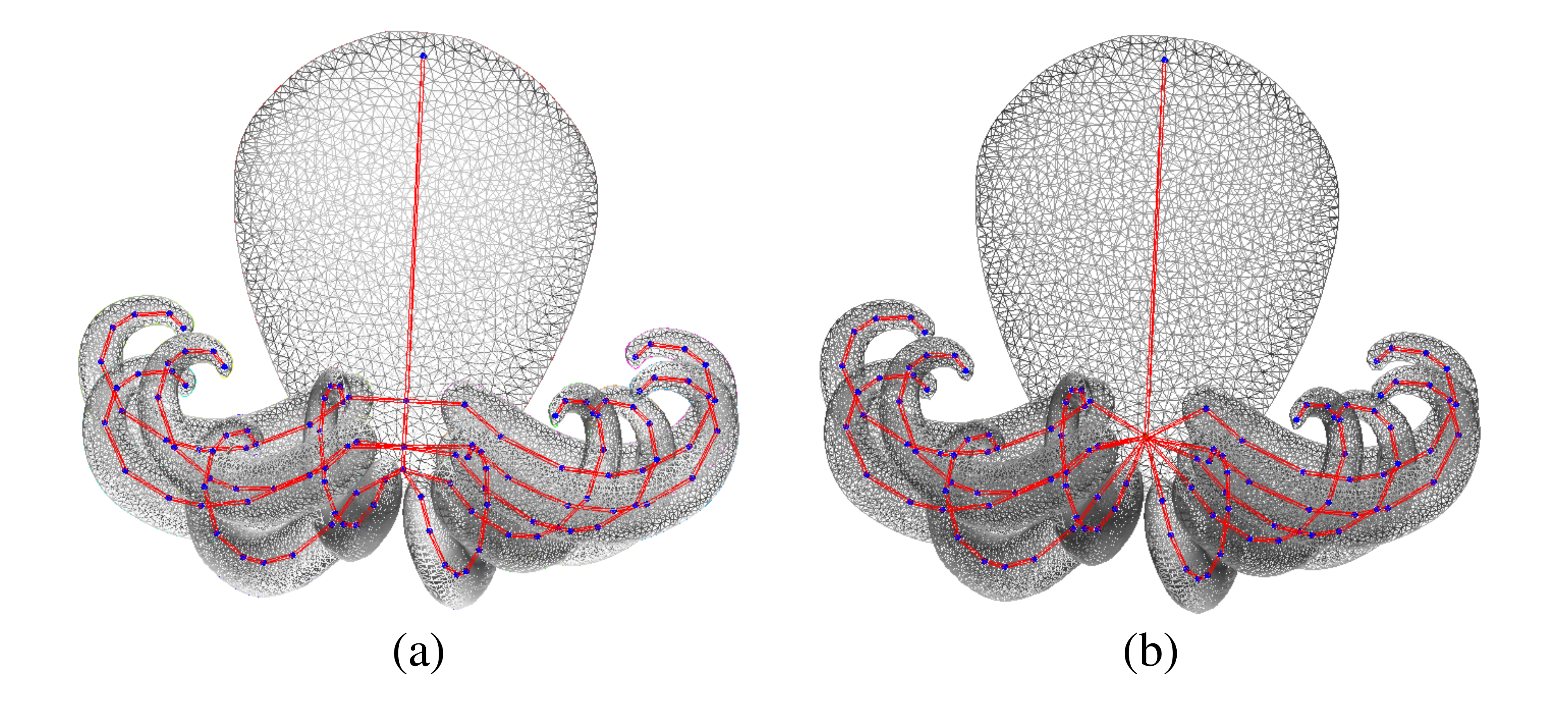}
	\caption{Merging. (a) Before merging. (b)After merging.}
	\label{fig:merging}
\end{figure}
Our global refinement method aims to generate a clean skeleton suitable for animation. Specifically, it solves two problems. Firstly, sub-skeleton connection usually induces redundant short bones, especially when two or more child-skeletons connect to similar positions on the parent-skeleton. Secondly, users need more control for skeleton complexity on different parts of the model, for example, users may expect a more complex skeleton for the hand than for the torso. To equip users with interaction easiness and flexibility. We introduce four operations and a multi-level strategy. The four operations are \emph{curve simplification}, \emph{joints merging}, \emph{branch pruning}, and \emph{edge collapsing}.  The multi-level complexity control contains three tiers: \emph{branch level}, \emph{subpart level}, and \emph{global level}. \emph{Simplifying} works on branch level, whereas \emph{merging}, \emph{pruning} and \emph{collapsing} work on subpart level and global level.
\paragraph{Branch level operation} 
We control branch complexity using BoundedDP of Section \ref{sec:local_skel_extract}. Specifically, we map the subpart-shape and the branch from 3D space into 2D space by rotating the 3D plane (Fig.~\ref{fig:intro}b, Fig.~\ref{fig:gallery} 1st row) back to XY plane, run the BoundedDP algorithm in 2D space, and then remap the simplified branch back to its original 3D position. The distance threshold for DP simplification is $\epsilon_{s}$, and the default value in our experiment is $\epsilon_{s}=5.0$.  Fig.~\ref{fig:refine_bird} (green boxes) shows simplifying the wing skeleton by increasing $\epsilon_{s}$.
\paragraph{Subpart and global level operation} 
$(i)$ \emph{Joints Merging}: We observe that junction joints usually cluster together after the user connects the child sub-skeleton to the parent sub-skeleton. The distance between any two junction nodes inside the cluster is small, thus it is reasonable to merge these junction nodes into one, see Fig.~\ref{fig:merging}. We perform a breadth-first search to find junction clusters constrained by a distance threshold, it is similar to the algorithm that finds a connected component of graph \cite{bfs_con_comp}. The distance threshold for merging is $\epsilon_{m}$, and we set its default value as $\epsilon_{m}=30.0$. $(ii)$ \emph{Branch Pruning}: When a JT-branch (the branch ending with a junction joint and a terminal joint) is shorter than a certain length, we remove it since it is insignificant to the overall structure. The distance threshold for trimming a JT-branch is $\epsilon_{t}$, and we set the default value as $\epsilon_{t}=30.0$. Yellow boxes of Fig.~\ref{fig:refine_bird} show such an example. $(iii)$ \emph{Edge Collapsing}: Junction joints and sleeve joints are internal joints. When a bone connecting two internal joints is short, it brings redundancy for an animatable skeleton, thus we collapse the two joints into one. The distance threshold for collapsing is $\epsilon_{c}$, and we set the default value as $\epsilon_{c}=10.0$. Black boxes of Fig.~\ref{fig:refine_bird}(b)-(c) show acquiring a more concise skeleton by increasing collapsing threshold $\epsilon_{c}$.
\paragraph{Multi-level complexity control}
We provide the user the option to control the skeleton structure by adjusting the above four parameters $\epsilon_{s}$, $\epsilon_{m}$, $\epsilon_{t}$, and $\epsilon_{c}$. Since a smaller threshold yields a more complex structure, the skeleton topology changes in a coarse-to-fine manner when users adjust the threshold. For example, a more complex hand skeleton can be acquired by setting a smaller value of $\epsilon_{s}$, more knuckles appear on the finger branch, shown in red boxes of Fig.~\ref{fig:refine_bird}a-b.
If the user needs the coarse representation, they can adjust the threshold to a higher value. Besides, users can select a single branch, a subpart-skeleton, or the global skeleton, and apply these operations on them. They can explore different skeleton structures simply by playing with these parameters.

\section{Results}
% TODO: \usepackage{graphicx} required
\begin{figure*}
	\centering
	\includegraphics[width=0.95\linewidth]{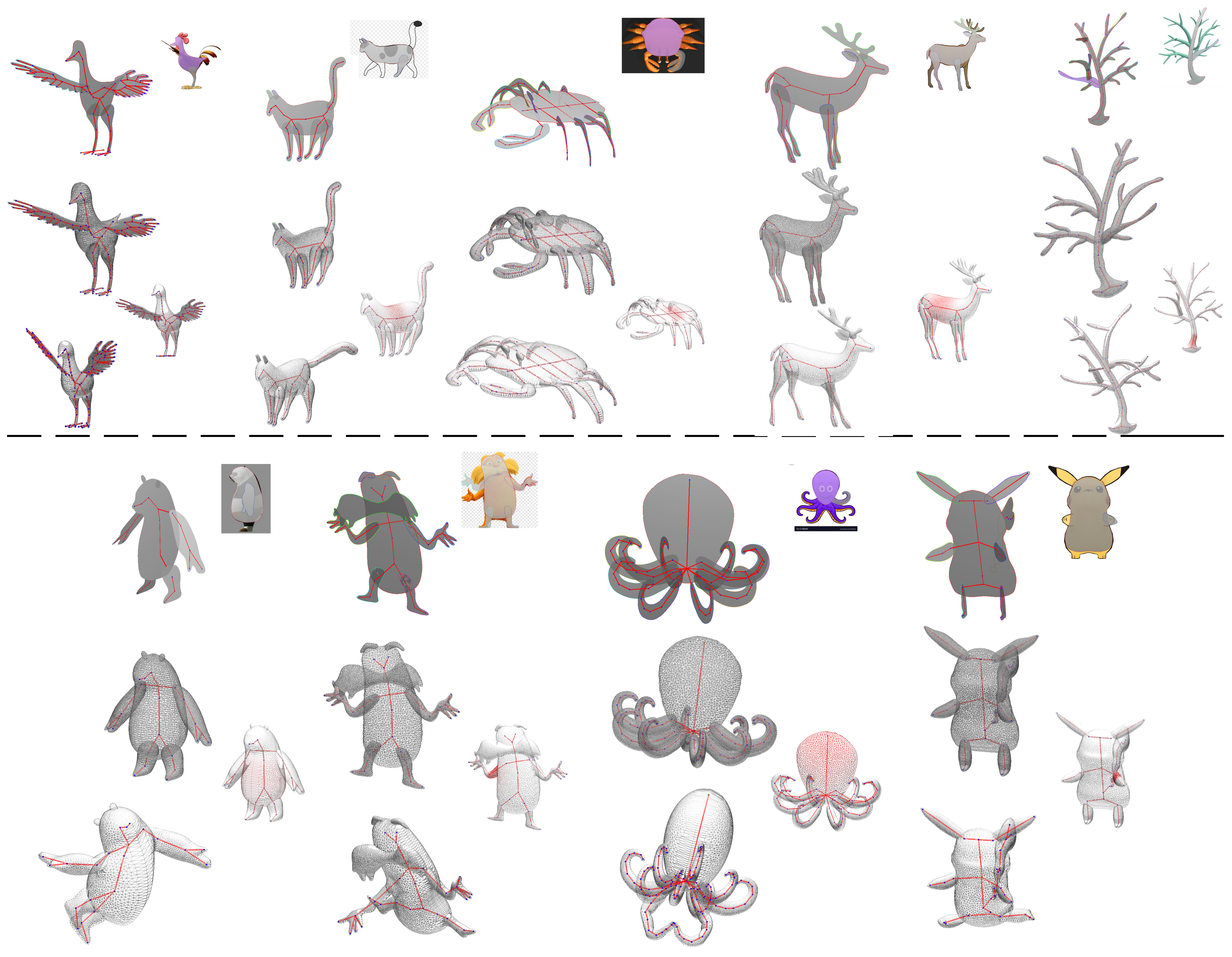}
	\caption{A gallery of models created by professionals using our system (zoom out the PDF to see the details). The first row shows the initial sketches from input image (right corner) and the silhouette polygons (drawn in 2D, then transformed to 3D). The second row shows the refined global skeleton along with the inflated 3D mesh. The third row shows the skinning weights (right corner) and the deformed models.} 
	\label{fig:gallery}
\end{figure*}
We implement our approach using C++, libIGL\footnote{https://libigl.github.io/}, and Qt. The current implementation, available at \url{https://github.com/jingma-git/RealSkel}, is tested on a AMD Ryzen 7 3700X 8-Core Processor 3.6GHz with 62GB RAM under Ubuntu20.04 LTS system. Our program runs at an interactive rate (see the accompanying video). Fig.~\ref{fig:gallery} shows a gallery of created models using our system.
\subsection{Implementation Details}
 We use CGAL's implementation to extract the straight skeleton \cite{sskel_impl_cgal_cacciola2004}, which runs in $O(n m+n \log n)$ time, where $n$ denotes the number of polygon vertices and $m$ denotes the reflex ones. For mesh creation, we first triangulate the polygon to get a 2D mesh \cite{triangle_shewchuk1996}, and generate the 3D mesh by an inflation algorithm  \cite{monster_mash_dvorovzvnak2020}. The height field of internal vertices is calculated by Poisson equation subject to the Dirichlet boundary condition and a constant height parameter $c$. We provide the user the option to change the model's thickness by adjusting the height parameter $c$. The sparse linear system is solved by LLT Cholesky factorization, which takes $O(n^3)$ time, where $n$ denotes the number of vertices in the 2D mesh. We also provide the user some utility functions such as sketch symmetry line, create symmetry part, rotation, translation, and scaling, etc. After all subparts and the global skeleton are constructed, we merge all parts into a watertight mesh based on the method from \cite{mesh_arrange_zhou2016}. To make our system fully automatic, we use the method from \cite{BBW} to compute the skinning weights for deformation, see  Fig.~\ref{fig:gallery} 3rd row. 
\subsection{Comparisons}
\label{sec:experiment}
%% Animation: root node selection
\begin{table*}
	\begin{center}
			\begin{tabular}{c|c|c|c|c|c|c|c|c|c|c|c|c}
			\toprule
			\multirow{3}{*}{Models} & \multicolumn{6}{c|}{Modeling Time(min)}& \multicolumn{6}{c}{Model Accuracy(1-5 stars)}                                                             \\ \cline{2-13} 
			& \multicolumn{2}{c|}{MonsterMash} & \multicolumn{2}{c|}{RigMesh} & \multicolumn{2}{c|}{\textbf{Ours.}} & \multicolumn{2}{c|}{MonsterMash} & \multicolumn{2}{c|}{RigMesh} & \multicolumn{2}{c}{\textbf{Ours.}}  \\ \cline{2-13} 
			       &Nov&Pro& Nov&Pro& Nov&Pro &  Nov&Pro& Nov&Pro& Nov&Pro                   \\ \midrule
			bird   &10 &6  & 40 &32 & 18 &14  &  2.5&3.0& 3.5&3.5& 4.5&4.5\\ \hline
			cat    &7  &4  & 29 &23 & 12 &10  &  3.0&3.0& 3.5&3.5& 4.5&5.0\\ \hline
            crab   &5.5&5  & 33 &28 & 17 &15  &  2.0&2.0& 3.0&3.0& 4.5&4.5\\ \hline
			deer   &5  &3.5& 42 &39 & 16 &13  &  2.5&3.0& 3.0&3.5& 4.5&4.5\\ \hline
			tree   &8  &6  & 28 &21 & 18 &12  &  3.0&3.5& 3.5&3.5& 4.0&4.5\\ \hline
			panda  &5  &3  & 30 &27 & 16 &11  &  1.5&1.5& 2.5&3.5& 4.5&4.5\\ \hline
			lorax  &3  &2.5& 21 &19 & 10 &8   &  3.5&3.5& 3.0&3.5& 4.5&5.0\\ \hline
			octopus&6.5&5  & 30 &22 & 19 &13  &  1.5&2.0& 3.0&3.0& 4.5&4.5\\ \hline
			pokemon&4  &2  & 32 &24 & 14 &10  &  2.0&2.0& 3.0&3.5& 4.5&4.5\\ \hline
			\textbf{Average}&6  &4.1&31.7&26.1&\textbf{15.6}&\textbf{11.8}& 2.4&2.6& 3.1&3.4& \textbf{4.4}& \textbf{4.6}\\
			\bottomrule
		\end{tabular}
	\end{center}
	\caption{Compare our system with MonsterMash and RigMesh. Nov: novices. Pro: professionals. Our system is easy for users to create high-quality models. The easiness is revealed by \emph{Modeling Time} and the quality is revealed by \emph{Model Accuracy}. The shorter the time is, the easier for the user to interact with the system. The more stars are given, the more accurate the model is. The modeling time of our system is half of the RigMesh, while the model accuracy of our system is higher than both MonsterMash and RigMesh.}\label{table:sys_comp}
\end{table*}

\begin{figure}
	\centering
	\includegraphics[width=0.95\linewidth]{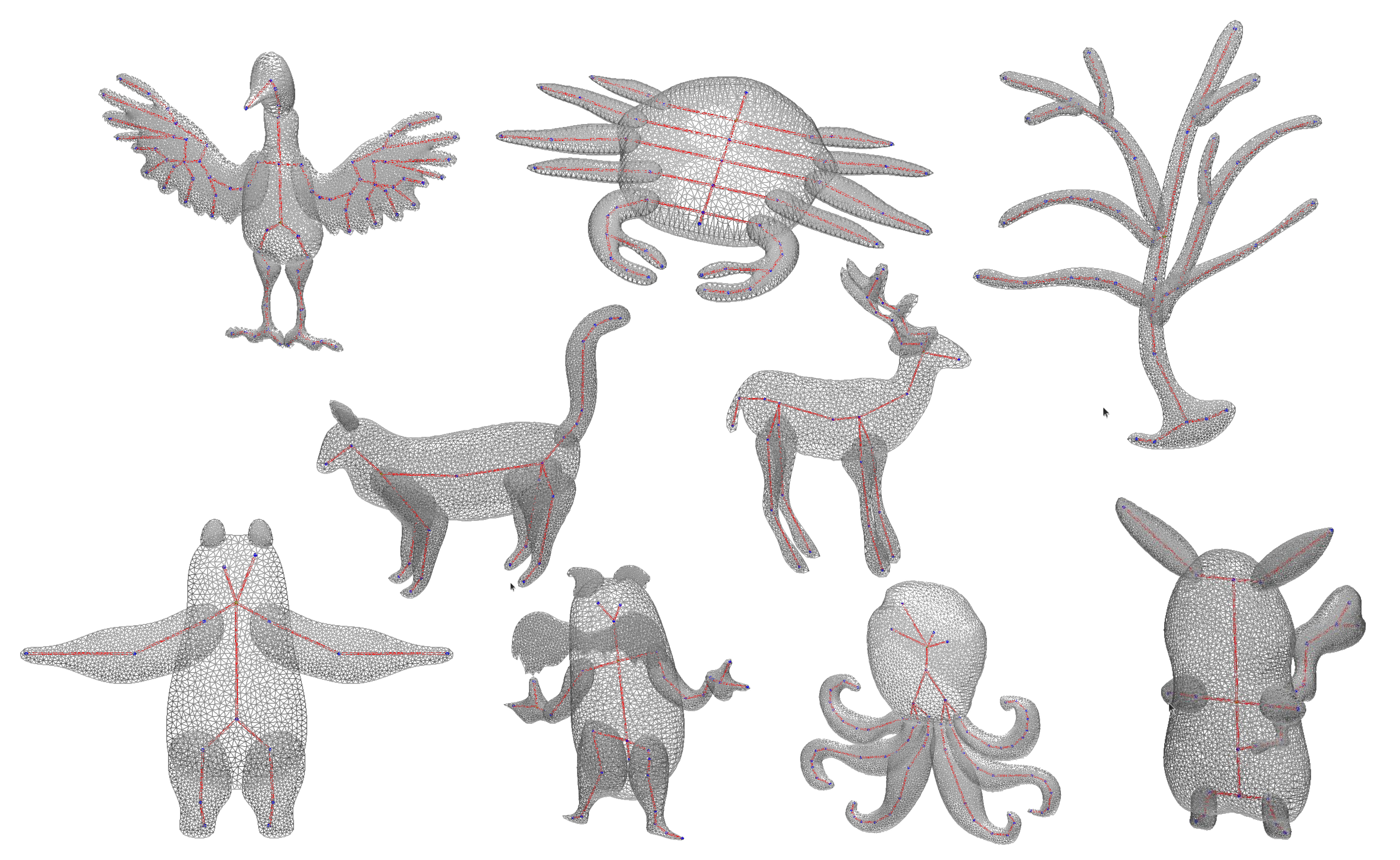}
	\caption{Models created by novices using our system. Models take an average 15 minutes to create.}
	\label{fig:novice} % ToDO: change to 9 models
\end{figure}
\begin{figure}
	\centering
	\includegraphics[width=0.95\linewidth]{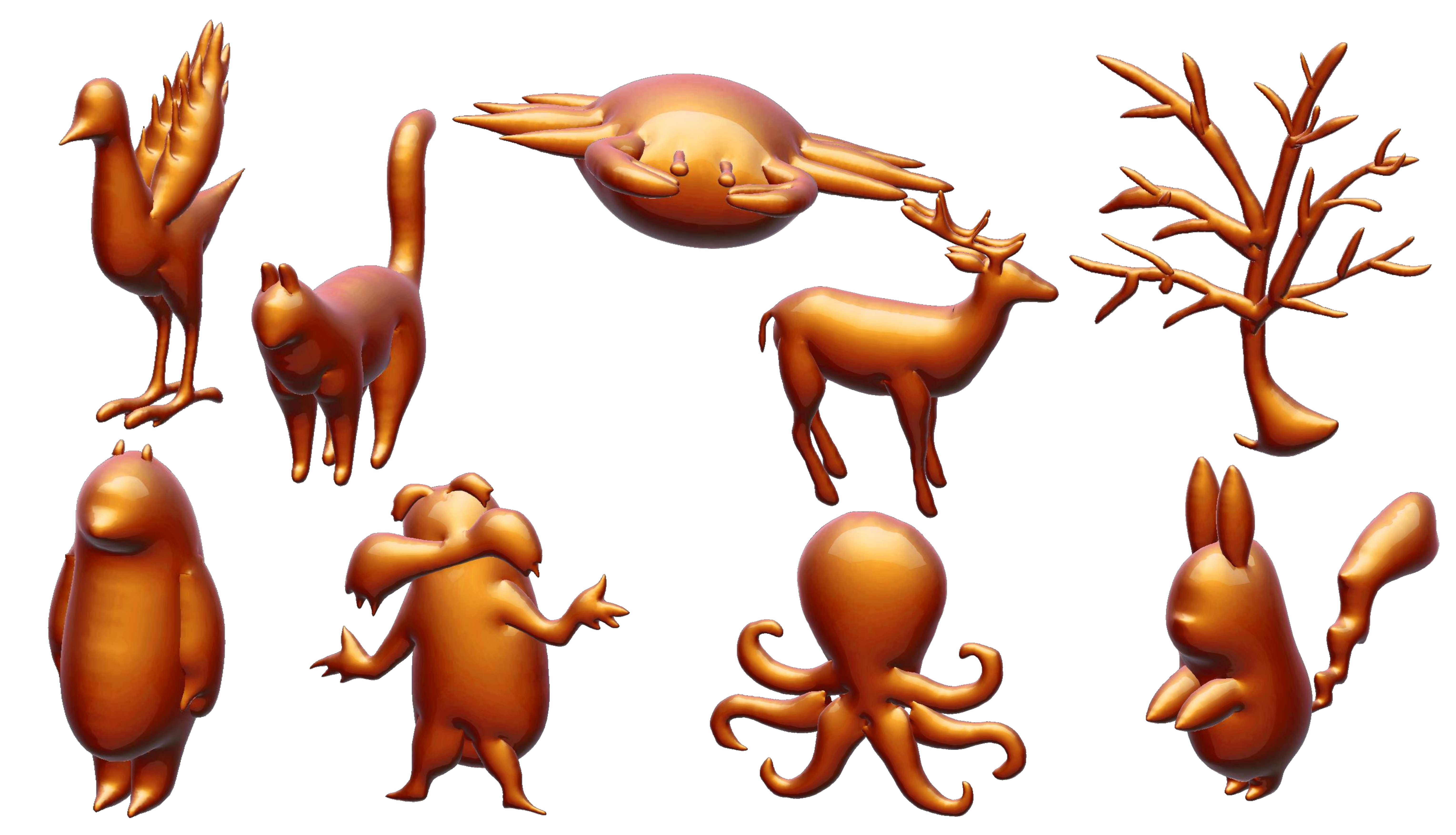}
	\caption{Models created using MonsterMash \cite{monster_mash_dvorovzvnak2020}. Models take an average 4 minutes to create. No skeleton is automatically generated. It only supports single-view modeling, which accelerates the modeling time but reduces the shape accuracy, see bird wings, cat ears, and crab body.}
	\label{fig:monstermash}
\end{figure}
\begin{figure}
	\centering
	\includegraphics[width=0.95\linewidth]{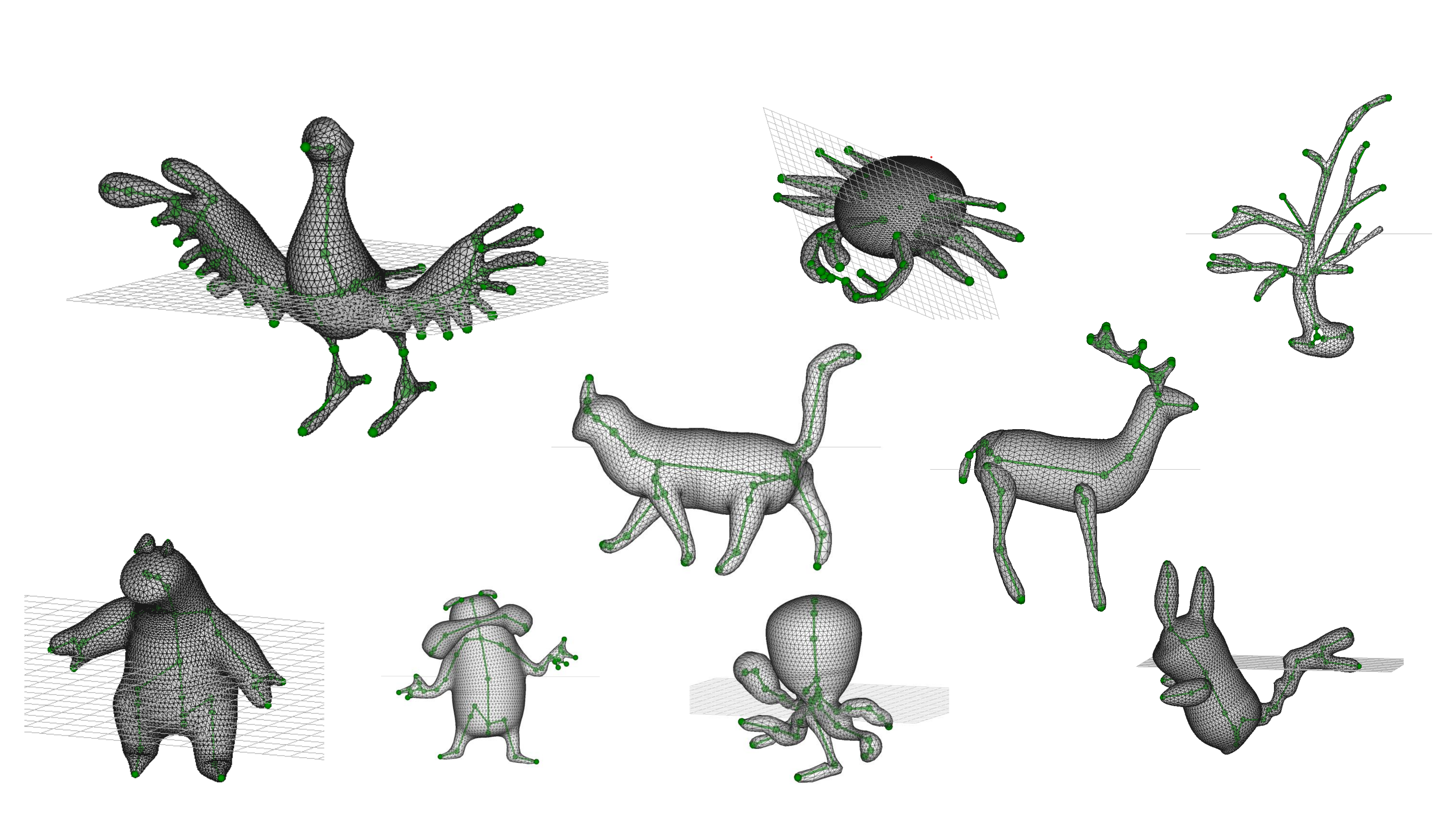}
	\caption{Models created using RigMesh \cite{rigmesh_borosan2012}. Models take an average 26 minutes to create. The skeleton joints are redundant and inaccurate, see cat butt, panda torso, and lorax shoulder.}
	\label{fig:rigmesh}
\end{figure}

We compare our system with MonsterMash\footnote{https://monstermash.zone/} and RigMesh\footnote{https://cragl.cs.gmu.edu/rigmesh/} to evaluate the interactive part of our system. Twelve users participate in the user study, three with modeling experience are assigned to the professional group, six without any modeling experience are assigned to the novice group, and the rest three are assigned to the jury group to judge the model accuracy (the quality of the mesh and the skeleton). We ask the first two groups to use MonserMash, RigMesh, and our system to create 3D models, and then show the models to the third group and ask them to fill a questionnaire regarding model accuracy. During the experiment, we give the participant a user manual, and record the time they take to be familiar with the system. On average, it takes eight minutes, twenty minutes, and twelve minutes to train the participants on MonsterMash, RigMesh, and our system respectively. After the user finishes training, they are allowed to proceed to the modeling stage.  The modeling time is shown in columns 2-7 of Table.~\ref{table:sys_comp}. Models created by professionals and novices using our system are shown in Fig.~\ref{fig:gallery} and Fig.~\ref{fig:novice} respectively. Models created by professionals using MonsterMash are shown in Fig.~\ref{fig:monstermash}. Models created by professionals using RigMesh are shown in Fig.~\ref{fig:rigmesh}. The model accuracy given by the jury group is shown in columns 8-13 of Table.~\ref{table:sys_comp} (MonsterMash models are accessed by mesh quality, RigMesh and our system are accessed by both mesh quality and skeleton quality). Compared with the other two systems,  our system takes less time for users to create accurate models.

\begin{figure}
	\centering
	\includegraphics[width=0.95\linewidth]{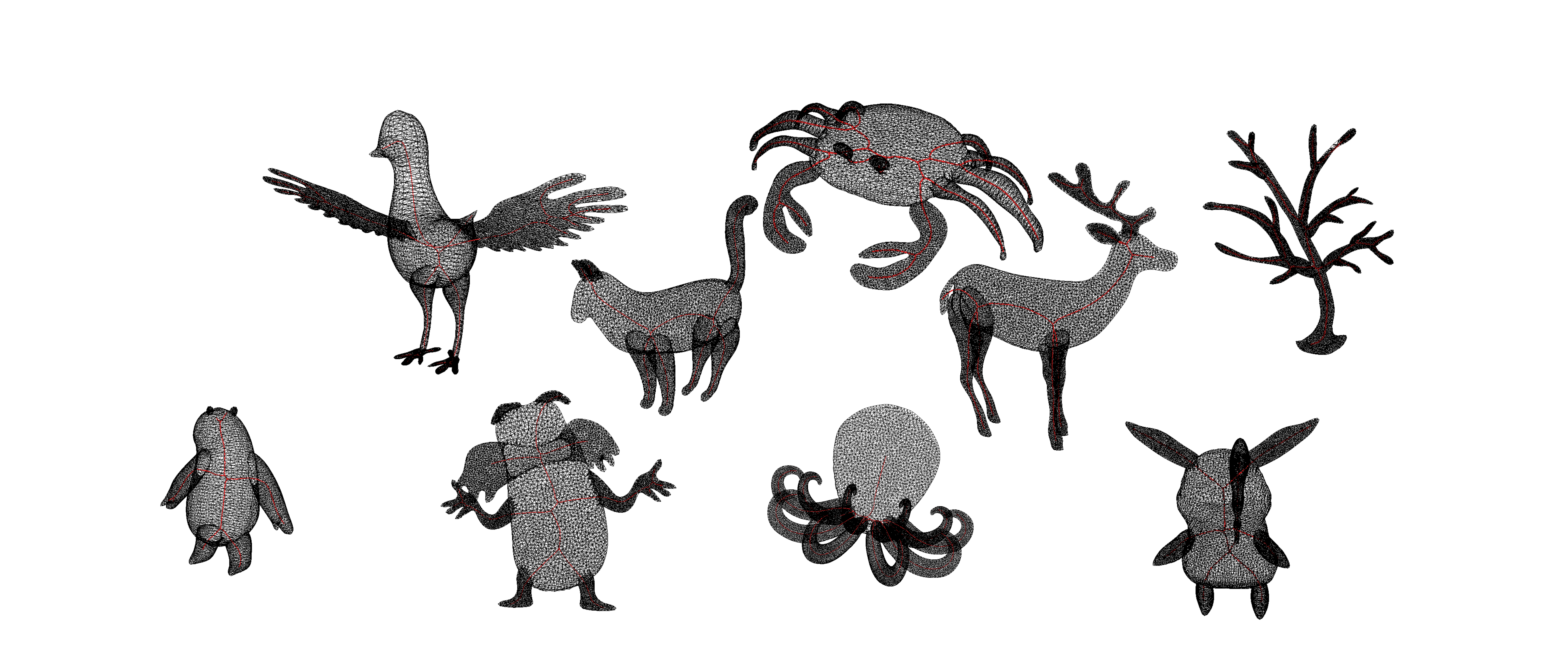}
	\caption{The curve skeleton extracted by MeanCurvature \cite{mean_curvature_tagliasacchi2012}. The skeleton is smooth and medially centered.}
	\label{fig:meancurvature}
\end{figure}

\begin{figure}
	\centering
	\includegraphics[width=0.95\linewidth]{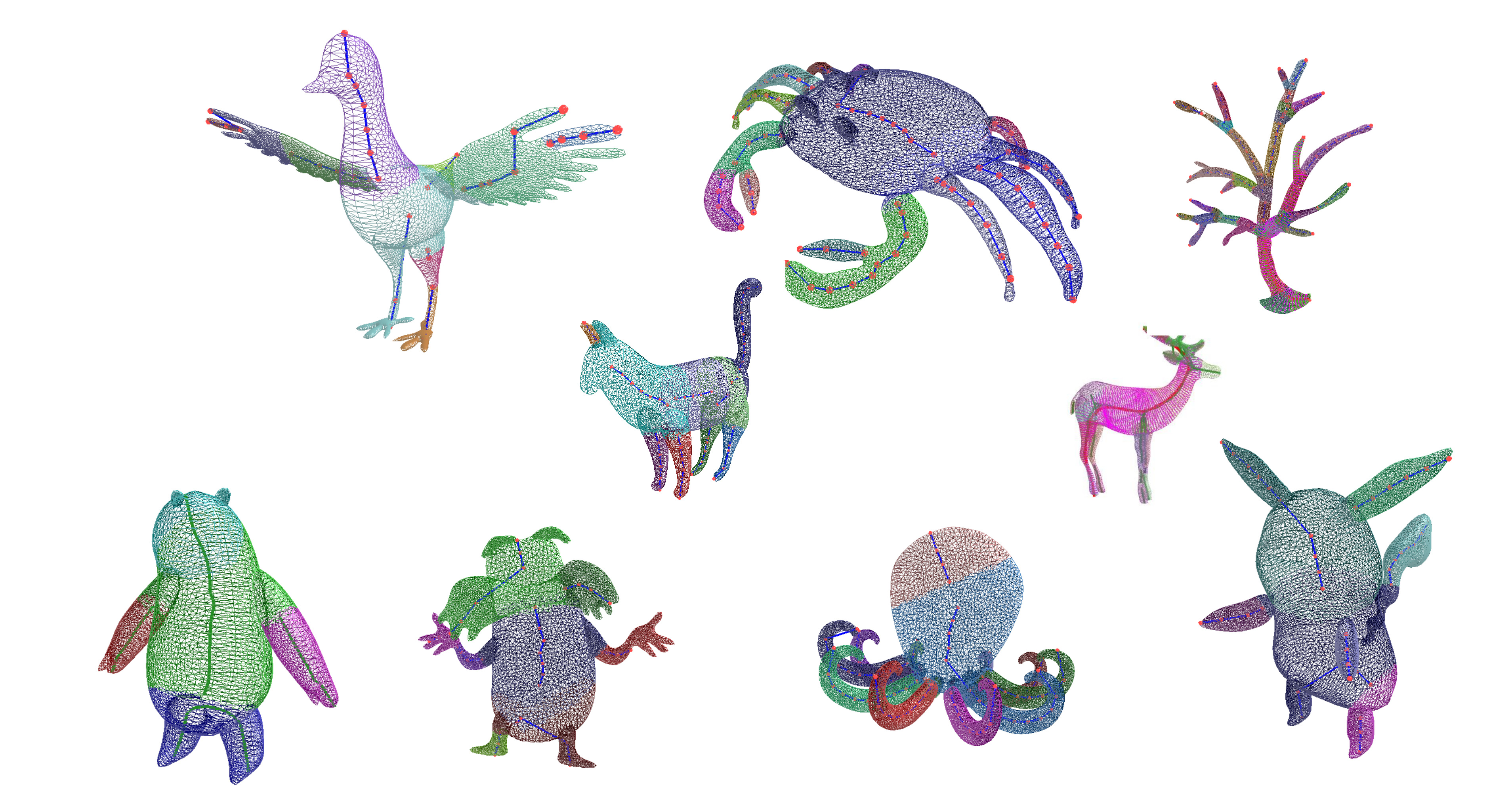}
	\caption{The disconnected curve skeleton extracted by GeneralCylinder \cite{general_cylinder_zhou2015}. A color patch represents a general cylinder. The skeleton is sensitive to the general cylinder decomposition, see bird wing, lorax head, and octopus body.}
	\label{fig:generalcylinder}
\end{figure}

\begin{table*}
	\begin{center}
		\begin{tabular}{c|c|c|c|c|c|c}
			\toprule
			\multirow{2}{*}{Models} &\multirow{2}{*}{\#vertices} & \multicolumn{2}{c|}{3D Skeletonization} & \multicolumn{3}{c}{2D Skeletonization}        \\ 
			\cline{3-4} \cline{5-7}
			& &MeanCurvature&GeneralCylinder&Zhang-Suen& ChordalAxisTransform& \textbf{Ours.} \\ \midrule
			bird    &10808&582 &22052&331&6&13\\ \hline
			cat	    &6980 &598 &18088&167&4&4 \\ \hline
			crab    &19079&860 &26382&412&6&10\\ \hline
			deer    &10091&745 &39587&217&4&10\\ \hline
			tree    &11920&681 &\textbf{43986}&141&4&7 \\ \hline
			panda   &5939 &406 &14057&254&6&5\\ \hline
			lorax   &8750 &719 &18135&212&6&9 \\ \hline
			octopus &17004&\textbf{1052}&21941&\textbf{560}&5&14\\ \hline
			pokemon &10487&672 &13195&345&6&7 \\ \hline
            \textbf{Average} &*  &701.8&24158.7&293.2&5.2&\textbf{8.8} \\
			\bottomrule              
		\end{tabular}
	\end{center}
	\caption{Execution time(\si{\milli\second}) of our algorithm compared with other algorithms. $\#vertices$ represents the number of vertices for merged mesh of Fig.~\ref{fig:gallery}. Columns 3-4 show the execution time of MeanCurvature and GeneralCylinder respectively. Columns 4-6 show the execution time of Zhang-Suen's thinning \cite{Zhang_thinning}, RigMesh ChordalAxisTransform \cite{rigmesh_borosan2012}, and our algorithm. In general, 2D skeletonization is faster than the 3D counterparts. Our method is significantly faster than MeanCurvature, GeneralCylinder and Zhang-Suen's Thinning, and it is slightly slower than ChordalAxisTransform. The relationship between $\#vertices$ and the algorithm's time complexity is revealed in Fig.~\ref{fig:vertices}.}
	\label{table:algo_time}
\end{table*}
We compare with other skeletonization methods to evaluate the algorithm part of our system. Our algorithm extracts skeleton from 2D polygon (sub-skeleton extraction), and then transforms the skeleton to the 3D space to construct the 3D skeleton (sub-skeleton connection and global refinement). Therefore, we compare our method with both 2D and 3D skeletonization methods. For 3D skeletonization algorithms, we choose  MeanCurvature \cite{mean_curvature_tagliasacchi2012} and GeneralCylinder \cite{general_cylinder_zhou2015} to compare.  MeanCurvature is based on mesh contraction, while GeneralCylinder extracts skeleton by slicing mesh to find ROSA \cite{rosa}. The merged mesh of Fig.~\ref{fig:gallery} is used as input to the 3D skeletonization algorithms. For 2D skeletonization algorithms, we choose Zhang-Suen's Thinning \cite{Zhang_thinning} and ChordalAxisTransform of RigMesh \cite{rigmesh_borosan2012} to compare. Zhang-Suen's algorithm is based on binary image thinning while ChordalAxisTransform extracts chordal axis from the constrained Delaunay triangulated polygon. We use the subpart contour of  Fig.~\ref{fig:gallery} (1st row) as the input to the 2D skeletonization algorithms. All experiments are conducted in the same machine.  Table~\ref{table:algo_time} compares the execution time of our algorithm with others. Our algorithm is significantly faster than MeanCurvature, GeneralCyliner, and ZhangSuen's Thinning. We also make qualitative comparisons. The skeleton quality of our algorithm is on par with MeanCurvature of Fig.~\ref{fig:meancurvature}, and is better than RigMesh of Fig.~\ref{fig:rigmesh} and GeneralCylinder of Fig.~\ref{fig:generalcylinder}.
% TODO: \usepackage{graphicx} required
\begin{figure}
	\centering
	\includegraphics[width=0.99\linewidth]{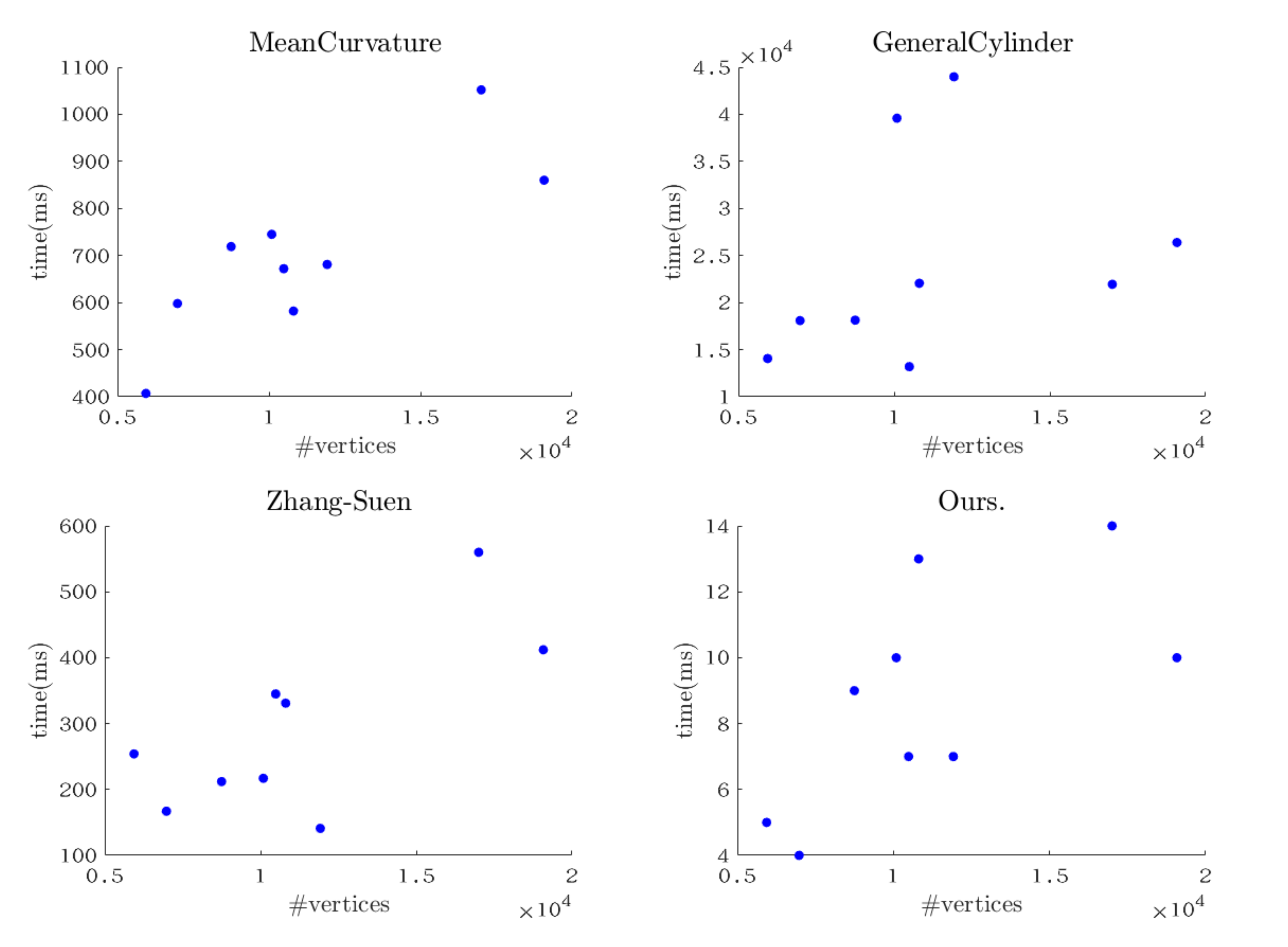}
	\caption{The \emph{execution time} is proportional to $\#vertices$.}
	\label{fig:vertices}
\end{figure}


\section{Conclusion and future work}
In this paper, we present an efficient framework to create the animatable skeleton in real-time for sketch modeling. We divide the big problem into three smaller ones: $(i)$ local sub-skeleton extraction; $(ii)$ sub-skeleton connection; $(iii)$ global skeleton refinement; and we conquer the problem one by one. To solve the first problem, we extract a straight skeleton by propagating the silhouette polygon inward, and we accelerate the algorithm by finding a concise shape approximation through DP simplification, a novel BoundedDP algorithm is then used to transfer the straight skeleton into the animatable bone skeleton. To solve the second problem of skeleton connection, we abstract the shape with general cylinders for bone and inscribed balls for joints, which enables a real-time intersection test among parts. And the precise attaching position is located by measuring Euclidean distance between parent bones and candidate child joints. To solve the third problem of creating different resolutions of skeletons, we propose four operations under three tiers of control. The branch-level operation allows users to control the bone complexity of a single axis. The subpart level and global level operation provide users with \emph{joints merging},  \emph{branch pruning}, and \emph{edge collapsing} operations to control the overall structure. The implemented system is demonstrated to be easy to use and enables complex shape  $\&$ rig creation. 

In the future, we intend to provide users more options to create shapes and skeletons by reusing existing 3d models. Specifically, our pipeline will not be limited to the immediately sketched shapes, users can also import 3D models created before and connect them to the newly sketched ones, and our system will automatically create an animatable skeleton for the model consisting of old and new shapes.
\section*{Acknowledgments}
This work is supported in part by the National Natural Science Foundation of China 
under 61972340, 61732015 and 51775492.

%\section*{References}

%%Vancouver style references.
\bibliographystyle{cag-num-names}
\bibliography{refs}
\newpage
\section*{Supplementary}
% Algorithm
% https://zhuanlan.zhihu.com/p/166418214
\IncMargin{1ex}
\begin{algorithm}
	%\SetAlgoNoLine
	\KwIn{A simple polygon $\mathcal{P}(\mathcal{V}, \mathcal{E})$ with vertices $\{v_i\}$ and edges $\{e_i\}$  oriented counter-clockwise.} 
	\KwOut{The straight skeleton $\mathcal{S}(\mathcal{P})$.}
	\SetKwFunction{InitDLAV}{InitDLAV}
	\SetKwFunction{CalBisector}{CalBisector}
	\SetKwFunction{CalBisectors}{CalBisectors}
	\SetKwFunction{CalIntersectTime}{CalIntersectTime}
	\SetKwFunction{CalIntersectPoint}{CalIntersectPoint}
	\SetKwFunction{FindOppositeEdgeInDLAV}{FindOppositeEdgeInDLAV}
	\SetKwFunction{CalSplitTime}{CalSplitTime}
	\SetKwFunction{CalSplitPoint}{CalSplitPoint}
	\SetKwFunction{CreateIntersectNode}{CreateIntersectNode}
	\SetKwFunction{UpdateDLAV}{UpdateDLAV}
	\SetKwFunction{CreateSplitNodes}{CreateSplitNodes}
	\BlankLine
	\tcp{Initialization}
	$L$ = \InitDLAV{$\mathcal{V}, \mathcal{E}$} \tcp*[r]{doubly linked active vertices} \label{algo:goto}
	\ForEach{$v_i$ in $L$}
	{
		$b_i$ = \CalBisector{$e_{i-1}$, $e_i$}\;
		update straight skeleton $\mathcal{S}(\mathcal{P})$ by $b_i$\;
	}
	\tcc{Compute initial events, place them in a priority queue $PQ$ ordered by offsetting time}
	\ForEach{$v_i$ in $L$}{
		\eIf{$v_i$ is convex}{
			$t_{isect}$ = \CalIntersectTime{$b_{i}$, $b_{i+1}$}\;
			$p_{isect}$ = \CalIntersectPoint{$b_{i}$, $b_{i+1}$}\;
			Add $EdgeEvent(b_i, b_{i+1}, p_{isect}, t_{isect})$ to $PQ$\;
			
		}{
			$e_{opp}$ = \FindOppositeEdgeInDLAV{$v_i$}\;
			$t_{split}$ = \CalSplitTime{$v_i$, $e_{opp}$}\;
			$p_{split}$ = \CalSplitPoint{$v_i$, $e_{opp}$}\;
			Add $SplitEvent(v_i, e_{opp}, p_{split}, t_{split})$ to $PQ$\;
		}
	}\label{algo:goto_e}
	
	\tcp{Propagation}
	\While{$PQ$ is not empty}{
		pop top $event$ from $PQ$\;
		\eIf{$event$ is an $EdgeEvent$}{
			\tcp{collapse edge}
			$v_{new}$ = \CreateIntersectNode{$EdgeEvent$}\;
			$b_{new}$ = \CalBisector{$EdgeEvent$}\; 
		}
		{
			\tcp{split polygon}
			$v_{new1}, v_{new2}$ = \CreateSplitNodes{$SplitEvent$}\;
			$b_{new1}, b_{new2}$ = \CalBisectors{$SplitEvent$}\; 
		}
		update DLAV $L$, straight skeleton $\mathcal{S}(\mathcal{P})$, and $PQ$ by new vertices and bisectors, see Line\ref{algo:goto}$\sim$\ref{algo:goto_e}\;
	}
	\caption{Straight skeleton extraction}
	\label{alg:one}
\end{algorithm}\DecMargin{1ex}

% TODO: \usepackage{graphicx} required

\end{document}